\documentclass[journal,article,submit,moreauthors,dvi2pdf]{Definitions/mdpi} 

\newtheorem{theo}{Theorem}

\newcommand{\argmin}{\mathop{\rm argmin}\limits}

\graphicspath{{./images/}}
\def\x{{\mathbf x}}

\def\y{{\mathbf y}}

\def\z{{\mathbf z}}
\def\v{{\mathbf v}}

\def\n{{\mathbf n}}

\def\u{{\mathbf u}}
\def\s{{\mathbf s}}

\def\d{{\mathbf d}}

\def\A{{\mathbf A}}

\def\D{{\mathbf D}}

\def\G{{\mathbf G}}

\def\R{{\mathbb R}}

\def\I{{\mathbf I}}
\def\P{{\mathbf P}}

\def\U{{\mathbf U}}
\def\V{{\mathbf V}}

\def\S{{\mathbf S}}

\def\PHI{{\mathbf \Phi}}

\def\Bmath{{\mathcal B}}

\def\Lmath{{\mathcal L}}

\DeclareMathOperator{\prox}{prox}

\DeclareMathOperator{\rank}{rank}
\DeclareMathOperator{\card}{card}

\DeclareMathOperator{\HTV}{HTV}
\DeclareMathOperator{\SSTV}{SSTV}
\DeclareMathOperator{\HSSTV}{HSSTV}
\DeclareMathOperator{\sgn}{sgn}

\firstpage{1} 
\pubvolume{xx}
\issuenum{1}
\articlenumber{5}
\pubyear{2020}
\copyrightyear{2020}
\history{Received: date; Accepted: date; Published: date}

\Title{A Constrained Convex Optimization Approach \\to Hyperspectral Image Restoration \\with Hybrid Spatio-Spectral Regularization}

\Author{Saori Takeyama $^{1,*}$\orcidA{}, Shunsuke Ono $^{1}$\orcidB{}, and Itsuo Kumazawa $^{1}$\orcidC{}}

\AuthorNames{Saori Takeyama, Shunsuke Ono and Itsuo Kumazawa}

\address{%
$^{1}$ \quad Tokyo Institute of Technology}

\corres{Correspondence: takeyama.s.aa@m.titech.ac.jp} 

\abstract{
We propose a new constrained optimization approach to hyperspectral (HS) image restoration.
Most existing methods restore a desirable HS image by solving some optimization problem, which consists of a regularization term(s) and a data-fidelity term(s). 
The methods have to handle a regularization term(s) and a data-fidelity term(s) simultaneously in one objective function, and so we need to carefully control the hyperparameter(s) that balances these terms. 
However, the setting of such hyperparameters is often a troublesome task because their suitable values depend strongly on the regularization terms adopted and the noise intensities on a given observation. 
Our proposed method is formulated as a convex optimization problem, where we utilize a novel hybrid regularization technique named Hybrid Spatio-Spectral Total Variation (HSSTV) and incorporate data-fidelity as hard constraints. 
HSSTV has a strong ability of noise and artifact removal while avoiding oversmoothing and spectral distortion, without combining other regularizations such as low-rank modeling-based ones. 
In addition, the constraint-type data-fidelity enables us to translate the hyperparameters that balance between regularization and data-fidelity to the upper bounds of the degree of data-fidelity that can be set in a much easier manner. 
We also develop an efficient algorithm based on the alternating direction method of multipliers (ADMM) to efficiently solve the optimization problem. 
Through comprehensive experiments, we illustrate the advantages of the proposed method over various HS image restoration methods including state-of-the-art ones.
}


\begin{document}


\section{Introduction}
Hyperspectral (HS) imagery has 1D spectral information including invisible light and narrow wavelength interval in addition to 2D spatial information and thus can visualize unseen intrinsic characteristics of scene objects and environmental lighting. 
This makes HS imaging a key technique in many applications in a wide range of fields, e.g., earth observation, agriculture, and medical and biological imaging~\cite{HSI1,HSI2,HSI3}.


Observed HS images are often affected by noise because of the small amount of light in narrow wavelength and/or sensor failure. 
Also, in compressive HS imaging scenarios~\cite{one-shotHSI1, one-shotHSI2}, we have to estimate a full HS image from a very small number of measurements. 
Thus, we need some methods for restoring desirable HS images from such degraded observations in HS applications.


\begin{table*}[t]
\begin{center}
  \caption{The feature of existing methods for HS image restoration.} 
  \label{existing_method_feature}
  \begin{tabular}{|c||c|c|c|c|} \hline
  \backslashbox{methods}{feature} & spatial correlation & spectral correlation & convexity & hyperparameters \\ \hline
  HTV \cite{HTV} & $\bigcirc$ & $\times$  & {\bf{convex}} & interdependent \\ 
  SSAHTV \cite{HTV} & $\bigcirc$ & $\bigtriangleup$  & {\bf{convex}} & interdependent \\ 
  SSTV \cite{SSTV} & $\bigtriangleup$  & $\bigcirc$ & {\bf{convex}} & interdependent \\ 
  ASSTV \cite{ASSTV_HSI} & $\bigcirc$ & $\bigcirc$ & {\bf{convex}} & interdependent \\ 
  LRM \cite{LRMR} & $\times$  & $\bigcirc$ & nonconvex & {\bf{independent}} \\
  LNWTV + LRM \cite{LRWTV, NRLRWTV} & $\bigcirc$ & $\bigcirc$ & {\bf{convex}} & interdependent \\ 
  HTV + LRM \cite{LRTV} & $\bigcirc$ & $\bigcirc$ & nonconvex & interdependent \\ 
  ASSTV + LRM \cite{ASSTV_HSI2, TWNNM} & $\bigcirc$ & $\bigcirc$ & nonconvex & interdependent \\ 
  SSTV + LRM \cite{sGLM, LRTDTV, LSSTV} & $\bigcirc$ & $\bigcirc$ & {\bf{convex}} & interdependent \\  
  SSTV + LRM \cite{SDTVLA, GLSSTV} & $\bigcirc$ & $\bigcirc$ & nonconvex & interdependent \\  
  {\bf{proposed}} & $\bigcirc$ & $\bigcirc$ & {\bf{convex}} & {\bf{independent}} \\ \hline 
  \end{tabular}
\end{center}
\end{table*}


Most HS image restoration methods are established based on optimization: a desirable HS image is characterized as a solution to some optimization problem, which consists of a regularization term and a data-fidelity term. 
The regularization term evaluates a-priori knowledge about underlying properties on HS images, and the data-fidelity term keeps the consistency with a given observation. 
Thanks to the design, these methods get a reasonable result under ill-posed or ill-conditioned scenarios typical in HS image restoration.

Regularization techniques for HS image restoration are roughly classified into two groups: total variation (TV)-based approach and low-rank modeling (LRM)-based one. 
TV models the total absolute magnitude of local differences to exploit the piecewise-smooth structures of an image. 
Many TV-based approaches~\cite{HTV, ASSTV_HSI, SSTV, LRWTV} have been proposed for HS image restoration. 
Besides, LRM-based approaches exploit the underlying low-rank structure in the spectral direction of an HS image. A popular example is the so-called Low-rank matrix recovery (LRMR)~\cite{LRMR}.

Many recent methods~\cite{LRWTV, NRLRWTV, LRTV, ASSTV_HSI2, sGLM, LRTDTV, LSSTV, SDTVLA, GLSSTV, TWNNM} combine TV-based and LRM-based approaches, and in general, they perform better than approaches using either regularization. 
This is because TV-based approaches model the spatial structure of an HS image whereas LRM-based approaches the spectral one. 
Naturally, the methods have to handle multiple regularization terms and a data-fidelity term(s) simultaneously in one objective function, and so the methods require to carefully control the hyperparameter(s) balancing these terms. 
Specifically, such hyperparameters are \textit{interdependent}, which means that a suitable value of a hyperparameter varies depending both on the multiple regularization terms used and the noise intensities on a given observation. 
Hence, the hyperparameter settings in such combined approaches are often troublesome tasks.
%
Table~\ref{existing_method_feature} summarizes the features of the methods reviewed in this section.

Based on the above discussion, we propose a new constrained convex optimization approach to HS image restoration. 
Our proposed method restores a desirable HS image by solving a convex optimization problem involving a new TV-based regularization and hard constraints on data-fidelity. 
The regularization, named Hybrid Spatio-Spectral Total Variation (HSSTV), is designed to evaluate two types of local differences: direct local spatial differences and local spatio-spectral differences in a unified manner to effectively exploit both the underlying spatial and spectral structures of an HS image. 
Thanks to this design, HSSTV has a strong ability of noise and artifact removal while avoiding oversmoothing and spectral distortion, without combining LRM. 
Moreover, the constrained-type data-fidelity in the proposed method enables us to translate interdependent hyperparameters to the upper bounds of the degree of data-fidelity that can be determined based only on the noise intensity. 
As a result, the proposed method has no interdependent hyperparameter. 
We also develop an efficient algorithm for solving the optimization problem based on the well-known alternating direction method of multipliers (ADMM)~\cite{ADMM1,DRS2,ADMM3,ADMMBoyd}.

The remainder of the paper is organized as follows. 
Section II introduces notation and mathematical ingredients. 
Section III reviews existing methods related to our method. 
In Section IV, we define HSSTV, formulate HS image restoration as a convex optimization problem involving HSSTV and hard-constraints on data-fidelity, and present an ADMM-based algorithm. 
Extensive experiments on denoising and compressed sensing (CS) reconstruction of HS images are given in Section V, where we illustrate the advantages of our method over several state-of-the-art methods. 
Section VI concludes the paper. 
The preliminary versions of this work, without mathematical details, deeper discussion, new applications, nor comprehensive experiments have appeared in conference proceedings~\cite{HSSTV, ICIP2019}.

\section{Preliminaries}\label{sec:P}
\subsection{Notation and Definitions}
In this paper, let $\R$ be the set of real numbers.
We shall use boldface lowercase and capital to represent vectors and matrices, respectively, and $:=$ to define something.
We denote the transpose of a vector/matrix by $(\cdot)^{\top}$, and the Euclidean norm (the $\ell_2$ norm) of a vector by $\|\cdot\|$. 

For notational convenience,
we treat an HS image ${\boldsymbol{\mathcal{U}}} \in \R^{N_v \times N_h \times B}$ as a vector $\u \in \R^{NB}$ ($N:=N_v N_h$ is the number of the pixels of each band, and $B$ is the number of the bands)
by stacking its columns on top of one another, i.e., the index of the component of the $i$th pixel in $k$th band is $i+(k-1)N$ (for $i=1,\ldots,N$ and $k = 1,\ldots, B$).

\subsection{Proximal Tools}\label{subsec:prox}
A function $f:\R^L\rightarrow(-\infty, \infty]$ is called \textit{proper lower semicontinuous convex} if
$\mbox{dom}(f) := \{\x\in\R^L |\;f(\x)<\infty\}\neq\emptyset$,
$\mbox{lev}_{\leq\alpha}(f):=\{\x \in\R^L|\; f(\x)\leq\alpha\}$ is closed for every $\alpha\in\R$,
and $f(\lambda\x+(1-\lambda)\y)\leq\lambda f(\x)+(1-\lambda)f(\y)$
for every $\x, \y\in\R^L$ and $\lambda\in(0,1)$, respectively.
Let $\Gamma_0(\R^L)$ be the set of all proper lower semicontinuous convex functions on $\R^L$.

The \textit{proximity operator}\cite{Moreau} plays a central role in convex optimization based on proximal splitting.
The proximity operator of $f \in \Gamma_0 (\R^L)$ with an index $\gamma > 0$ is defined by
\begin{equation*} 
\prox_{\gamma f} (\x) := \argmin_{\y} f(\y) + \frac{1}{2 \gamma} \| \y - \x\|^2.
\end{equation*}

We introduce the indicator function of a nonempty closed convex set $C \subset \R^L$, which is defined as follows:
\begin{align}
\iota_{C}(\x) := \left\{
\begin{array}{l l}
0, &\mbox{if} ~ \x \in C, \nonumber \\
\infty , &\mbox{otherwise}. \nonumber \\
\end{array}
\right.
\end{align}
Then, for any $\gamma>0$, its proximity operator is given by
\begin{equation*} 
\prox_{\gamma \iota_{C}}(\x) = P_C(\x) :=\argmin_{\y \in C} \|\x-\y\|,
\end{equation*}
where $P_C(\x)$ is the metric projection onto $C$.

%

\subsection{Alternating Direction Method of Multipliers (ADMM)}\label{subsec:ADMM}
ADMM \cite{ADMM1,DRS2,ADMM3,ADMMBoyd} is a popular proximal splitting method, and it can solve convex optimization problems of the form:
\begin{equation}\label{ADMMequation}
\min_{\x,\z} f(\x) + g(\z) ~\mbox{s.t.}~ \z=\G\x,
\end{equation}
where $f\in\Gamma_0(\R^{L_1})$, $g\in\Gamma_0(\R^{L_2})$, and $\G\in\R^{L_2 \times L_1}$.
Here, we assume that $f$ is quadratic, $g$ is \textit{proximable}, i.e., the proximity operator of $g$ is computable in an efficient manner, and $\G$ is a full-column rank matrix.
For arbitrarily chosen $\z^{(0)}, \d^{(0)}$ and a step size $\gamma>0$,
ADMM iterates the following steps:
\begin{align}
\left \lfloor
\begin{array}{l}\label{ADMM_renew}
\x^{(n+1)} = \argmin_{\x} f(\x) + \frac{1}{2\gamma}\|\z^{(n)} - \G\x -\d^{(n)}\|^2, \\
\z^{(n+1)} = \mbox{prox}_{\gamma g}(\G\x^{(n+1)} + \d^{(n)}), \\
\d^{(n+1)} = \d^{(n)} + \G\x^{(n+1)} - \z^{(n+1)},
\end{array}
\right.
\end{align}

Convergence property of ADMM is given as follows.
\begin{theo}[Convergence of ADMM \cite{ADMM3}]\label{ADMMconv}
  Consider Prob.~\eqref{ADMMequation}, and assume that $\G^\top\G$ is invertible
  and that a saddle point of its unaugmented Lagrangian
  $\Lmath_0(\x,\z,\y):=f(\x)+g(\z)-\langle\d, \G\x-\z\rangle$
  exists.\footnote{
    A triplet $(\hat\x,\hat\z,\hat\d)$ is a saddle point of an unaugmented Lagrangian
    $\Lmath_0$ if and only if
    $\Lmath_0(\hat\x, \hat\z, \d)\leq\Lmath_0(\hat\x, \hat\z, \hat\d)\leq\Lmath_0(\x, \z,\hat\d)$,
    for any $(\x,\z,\d)\in\R^{L_1}\times\R^{L_2}\times\R^{L_2}$.}
  Then the sequence $(\x_n)_{n>0}$ generated by \eqref{ADMM_renew} converges to an optimal solution to Prob.~\eqref{ADMMequation}.
\end{theo}

\section{Related Works}\label{sec:RW}
In this section, we elaborate on existing HS image restoration methods based on optimization.

\subsection{TV-based Methods} \label{subsec:TV}
The methods proposed in \cite{HTV, SSTV, ASSTV_HSI} restore a desirable HS image by solving a convex optimization problem involving TV-based regularization.
Let $\bar{\u} \in \R^{NB}$ be the desirable HS image, and the authors assume that an observation $\v \in \R^{NB}$ is modeled as follows:
\begin{equation*}
\v = \bar{\u} + \s + \n,
\end{equation*}
where $\n$ and $\s$ are an additive white Gaussian noise and a sparse noise, respectively.
Here, the sparse noise corrupts only a few pixels in the HS image but heavily, e.g., impulse noise, salt-and-pepper noise, and line noise.
The observation and the restoration problem of the methods are given by the following forms:
\begin{equation}\label{prob:TV_based}
\min_{\u,\s} \|\v-\u-\s\|^2 + \lambda_1\mathcal{R}_{\mathrm{TV}}(\mathbf{u}) + \lambda_2\|\s\|_1,
\end{equation}
where $\mathcal{R}_{\mathrm{TV}}$ is a regularization function based on TV, and $\lambda_1$ and $\lambda_2$ are hyperparameters.
Here, The first and third terms evaluate data-fidelity on Gaussian and sparse noise, respectively.
The hyperparameters $\lambda_1$ and $\lambda_2$ represent the priorities of each term.
If we can choose suitable values of the hyperparameters, then this formulation yields high-quality restoration.
However, the hyperparameters are interdependent, which means that suitable values of the hyperparameters vary depending on the used TV-based regularization term and the noise intensities on a given observation.
Therefore, the settings of the hyperparameters are a very important but troublesome task.

\begin{figure}[t]
\begin{center}
 \includegraphics[bb= 0 0 771 636, width=0.6\hsize]{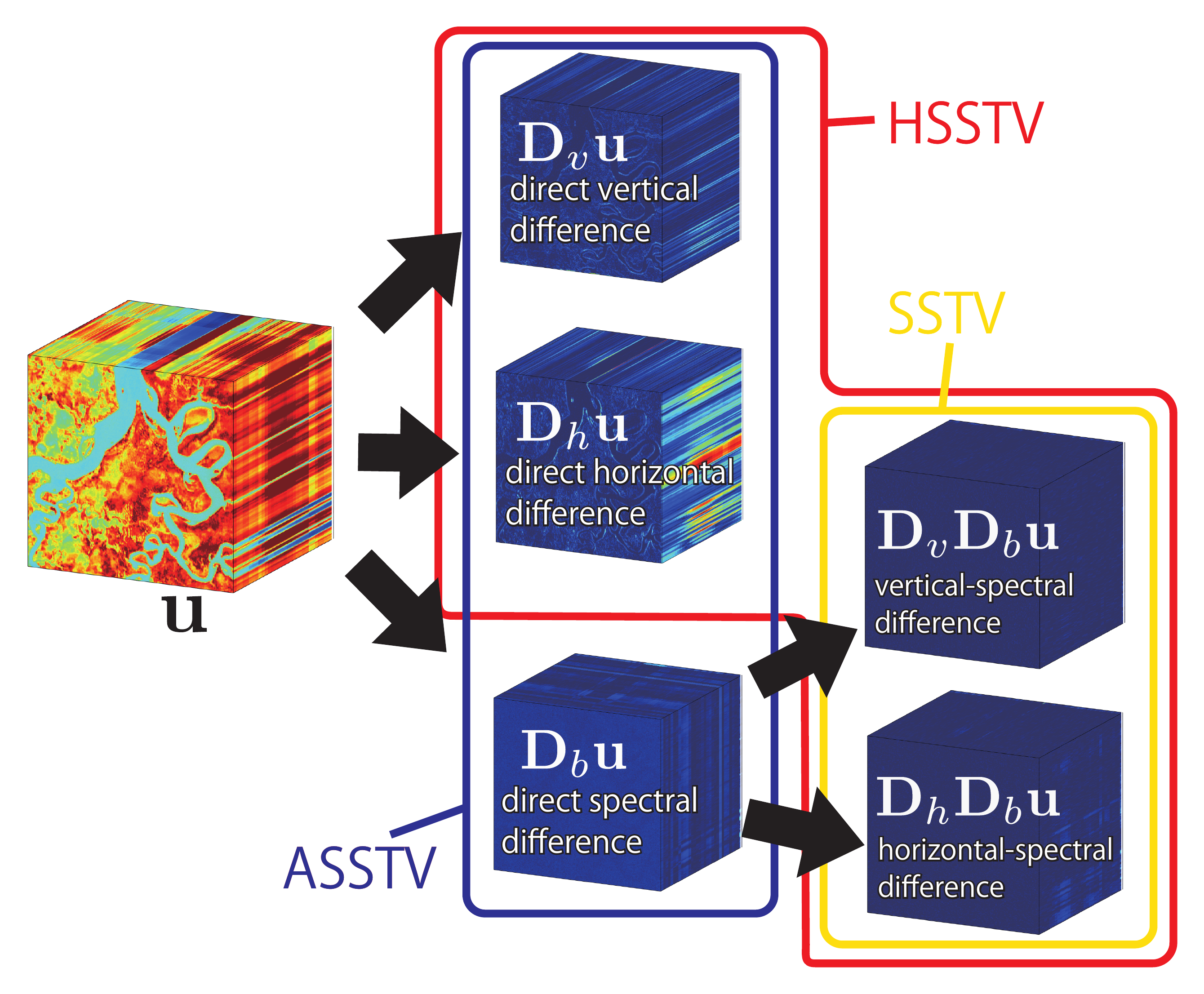}
 \caption{Calculation of local differences in SSTV, ASSTV and our HSSTV.
 SSTV evaluates the $\ell_1$ norm of spatio-spectral differences (yellow line).
 ASSTV evaluates the $\ell_1$ norm of direct spatial and spectral differences (blue line).
 HSSTV evaluates the mixed $\ell_{1,p}$ norm
 of both direct spatial and spatio-spectral differences (red line).}
 \label{HSSTV_image}
\end{center}
\end{figure}

In the following, we explain each TV. 
Let $\D = (\D_v^{\top} \D_h^{\top})^{\top} \in \R^{2NB \times NB}$ be spatial differences operator with $\D_v$ and $\D_h$ being vertical and horizontal differences operator, respectively, and spectral differences operator are $\D_b \in \R^{NB \times NB}$.
In \cite{HTV, SSTV, ASSTV_HSI}, HTV, ASSTV, and SSTV are defined as follows: 
\begin{align}
\HTV(\u) &:= \|\D\u\|_{\mathrm{TV}}, \label{def_HTV}\\
\mathrm{ASSTV}(\u) &:= \tau_v \|\D_v\u\|_1 + \tau_h \|\D_h\u\|_1 + \tau_b \|\D_b\u\|_1, \label{def_ASSTV}\\
\SSTV(\u) &:= \|\D\D_b \u\|_1, \label{def_SSTV}
\end{align}
where $\|\cdot\|_{\mathrm{TV}}$ is a TV norm, which takes the $\ell_{2}$ norm of spacial difference vectors for all band and then summing up for all spatial pixels, and $\tau_v$, $\tau_h$, and $\tau_b$ are the weight of the vertical, horizontal, and spectral differences.
HTV evaluates direct spatial piecewise-smoothness and can be seen as a generalization of the standard color TV~\cite{VTVBresson2008}. 
HTV does not consider spectral correlation, resulting in spatial oversmoothing.
To consider spectral correlation, the authors of \cite{HTV} proposed SSAHTV.
SSAHTV is a weighted HTV, and the weight is determined by spectral information.
However, since SSAHTV does not directly evaluate spectral correlation, it still causes spatial oversmoothing.
ASSTV evaluates direct spatial and spectral piecewise-smoothness (Fig.~\ref{HSSTV_image}, blue line).
The weights $\tau_v$, $\tau_h$, and $\tau_b$ in \eqref{def_ASSTV} balance the smoothness related to vertical, horizontal, and spectral differences, respectively.
Owing to the definition, ASSTV can evaluate spatial and spectral correlation, but it produces spectral oversmoothing even if we carefully adjust $\tau_v$, $\tau_h$, and $\tau_b$.
SSTV evaluate a-prior knowledge on HS images using spatio-spectral piecewise-smoothness.
It is derived by calculating spatial differences through spectral differences (Fig.~\ref{HSSTV_image}, yellow line).
SSTV can restore a desirable HS image without any weight, but it produces noise-like artifacts especially when a given observation is contaminated by heavy noise and/or degradation.

\subsection{LRM-based Method}
LRMR~\cite{LRMR} is one of the popular LRM-based methods for HS image restoration, which evaluates the low rankness of an HS image in the spectral direction.
To preserve the local details, LRMR restores a desirable HS image through patch-wise processing.
Each patch is a local cube of the size of $q \times q \times B$, and LRMR handles it as a matrix of size $q^2 \times B$ that is obtained by lexicographically arranging the spatial vectors in the patch cube in the row direction.
The observation model is expressed 
like Sec.~\ref{subsec:TV}, and the restoration problem is formulated as follows:
\begin{equation}
\min_{\U_{i,j}, \S_{i,j}} \|\V_{i,j} - \U_{i,j} - \S_{i,j}\|_{F}^2 ~\mathrm{s.t.}~ \rank(\U_{i,j})\leq r,~ \card(\S_{i,j}) \leq k, \label{prob:LRMR}
\end{equation}
where $\U_{i,j}$, $\V_{i,j}$, and $\S_{i,j}$ represents the patches of a restored HS image, an observation, and a sparse noise, respectively, which are centered at (i,j) pixel.
Then, The $\|\cdot\|_{F}$ is a Frobenius norm, $\rank(\cdot)$ represents a rank function, and $\card(\cdot)$ is a cardinality function.
The method evaluates the low rankness of the estimated HS image and sparsity of the sparse noise by limiting the number of the rank of $\U_{i,j}$ and the cardinality of $\S_{i,j}$ using $r$ and $k$, respectively.
Thanks to the design, LRMR achieves high-quality restoration for especially spectral information.
Meanwhile, since LRMR does not fully consider spatial correlation, the result by LRMR tends to have spatial artifacts when an observation is corrupted by heavy noise and/or degradation.
Besides, the rank and cardinality functions are non-convex, and so it is a troublesome task to seek the global optimal solution of Prob.~\eqref{prob:LRMR}.

\subsection{Combined Method}
The methods \cite{LRWTV, NRLRWTV, LRTV, ASSTV_HSI2, sGLM, LRTDTV, LSSTV, SDTVLA, GLSSTV, TWNNM} combine TV-based and LRM-based approaches. 
Since they can evaluate multiple types of a-priori knowledge, i.e., piecewise-smoothness and low rankness, they can restore a more desirable HS image than the approaches only using TV-based or LRM-based regularization.
Besides, some methods \cite{LRWTV, NRLRWTV, sGLM, LRTDTV, LSSTV, TWNNM} approximate the rank and cardinality functions by their convex surrogates.
As a result, the restoration problems are convex and can be solved by optimization methods based on proximal splitting. 

However, the methods have to handle multiple regularization terms and/or a data-fidelity term(s) simultaneously in one objective function, and so they require to carefully control the hyperparameters balancing these terms.
Since the hyperparameters rely on both the regularizations and the noise intensity on an observation, i.e., the hyperparameters are interdependent, the hyperparameter settings are often troublesome tasks. 

\section{Proposed Method}\label{sec:PM}
\subsection{Hybrid Spatio Spectral Total Variation}\label{subsec:HSSTV}
We propose a new regularization technique for HS image restoration, named HSSTV.
HSSTV simultaneously handles both direct local spatial differences and local spatio-spectral differences of an HS image. 
Then, HSSTV is defined by
\begin{equation}\label{HSSTV}
\HSSTV(\u) := \| \A_{\omega} \u \|_{1,p} \mbox{ with }
\A_{\omega} := \left(
\begin{array}{c}
\D\D_b\\
\omega\D\\
\end{array}
\right),
\end{equation}
where $\|\cdot\|_{1,p}$ is the mixed $\ell_{1,p}$ norm, and $\omega \geq 0$.
We assume $p = 1$ or $2$, i.e., the $\ell_1$ norm ($\|\cdot\|_{1,1} = \|\cdot\|_1$) or the mixed $\ell_{1,2}$ norm, respectively.
We would like to remark that we can also see $\ell_1$-HSSTV ($p = 1$) as \textit{anisotropic} HSSTV and $\ell_{1,2}$-HSSTV ($p = 2$) as \textit{isotropic} HSSTV.

In \eqref{HSSTV}, $\D\D_b\u$ and $\D\u$ correspond to local spatio-spectral and direct local spatial differences, respectively, as shown in Fig.~\ref{HSSTV_image} (red lines).
The weight $\omega$ adjusts the relative importance of direct spatial piecewise-smoothness to spatio-spectral piecewise-smoothness.
HSSTV evaluates two kinds of smoothness by taking the $\ell_p$ norm ($p=1$ or $2$) of these differences associated with each pixel and then summing up for all pixels, i.e., calculating the $\ell_1$ norm.
Thus, it can be defined via the mixed $\ell_{1,p}$ norm.
When we set $\omega = 0$ and $p = 1$, HSSTV recovers SSTV as \eqref{def_SSTV}, meaning that HSSTV can be seen as a generalization of SSTV.

\begin{figure}[t]
\begin{center}
\begin{minipage}[t]{0.24\hsize}
\includegraphics[bb = 0 0 256 256, width=\hsize]{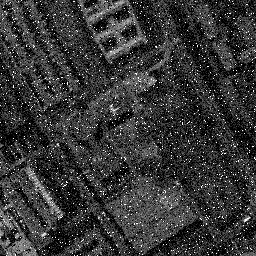}
\end{minipage}
\begin{minipage}[t]{0.24\hsize}
\includegraphics[bb = 0 0 256 256, width=\hsize]{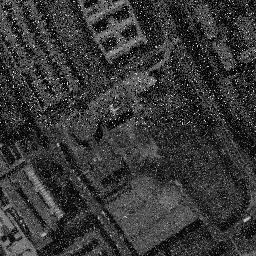}
\end{minipage}
\begin{minipage}[t]{0.24\hsize}
\includegraphics[bb = 0 0 256 256, width=\hsize]{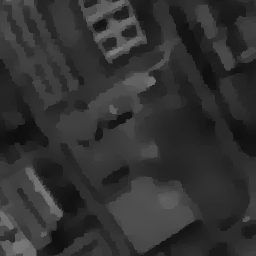}
\end{minipage}
\begin{minipage}[t]{0.24\hsize}
\includegraphics[bb = 0 0 256 256, width=\hsize]{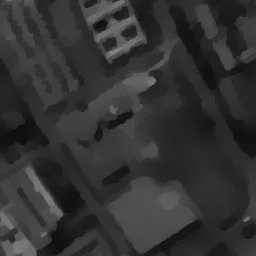}
\end{minipage}

\begin{minipage}[t]{0.24\hsize}
\centerline{\footnotesize{observation}}
\end{minipage}
\begin{minipage}[t]{0.24\hsize}
\centerline{\footnotesize{SSTV}}
\end{minipage}
\begin{minipage}[t]{0.24\hsize}
\centerline{\footnotesize{$\ell_1$-HSSTV}}
\end{minipage}
\begin{minipage}[t]{0.24\hsize}
\centerline{\footnotesize{$\ell_{1,2}$-HSSTV}}
\end{minipage}

\caption{Restored HS images from an observation contaminated by similar noise in adjacent bands (the upper half area) and random noise (the lower half).}
\label{fig:adjacent_band_noise}
\end{center}
\end{figure}

As reviewed in Sec.~\ref{sec:RW}, since SSTV only evaluates spatio-spectral piecewise-smoothness, it cannot remove similar noise in adjacent bands.
The direct spatial differences in HSSTV help to remove such noise.
Fig.~\ref{fig:adjacent_band_noise} is restored HS images from an observation contaminated by similar noise in adjacent bands (the upper half area) and random noise (the lower half area).
One can see that large noise remains in the upper half area of the result by SSTV.
In contrast, HSSTV effectively removes all noise.
However, since minimizing the direct spatial differences strongly promotes spatial piecewise-smoothness, HSSTV produces spatial oversmoothing when the weight $\omega$ is large.
Thus, the weight $\omega$ should be set to less than one, as will be demonstrated in Sec.~\ref{sec:E}.

\subsection{HS Image Restoration by HSSTV}\label{subsec:restoration}
We consider to restore a desirable HS image $\bar{\u} \in \R^{NB}$ from an observation $\v \in \R^{M} ~(M \leq NB)$ contaminated by a Gaussian-sparse mixed noise. 
The observation model is given by the following form: 
\begin{equation}\label{model}
 \v=\PHI \bar{\u} + \n + \s,
\end{equation}
where $\PHI \in \R^{M \times NB}$ is a matrix representing a linear observation process, e.g., random sampling, $\n \in \R^{M}$ is a Gaussian noise with the standard deviation $\sigma$,  and $\s \in \R^{M}$ is a sparse noise.

Based on the above model, we formulate HS image restoration using HSSTV as the following optimization problem:
\begin{align}\label{prob:HSSTV_Gaussian}
&\min_{\u, \s} \HSSTV(\u) \nonumber \\
& \mbox{ s.t. }
\left [ 
\begin{array}{l}
\PHI\u + \s \in \Bmath^{\v}_{2,\varepsilon} := \{ \x \in \R^{M} | \|\v - \x\| \leq \varepsilon\}, \\
\s \in \Bmath_{1, \eta} := \{ \x \in \R^{M} | \|\x\|_{1} \leq \eta \}, \\
\u \in [\mu_{\min}, \mu_{\max}]^{NB},
\end{array}
\right.
\end{align}
where $\Bmath_{2,\varepsilon}^{\v}$ is a $\v$-centered $\ell_2$-norm ball with the radius $\varepsilon > 0$, $\Bmath_{1, \eta}$ is a $\mathbf{0}$-centered $\ell_1$-norm ball with the radius $\eta > 0$, and $[\mu_{\min}, \mu_{\max}]^{NB}$ is a dynamic range of an HS image ($\mu_{\min} < \mu_{\max}$). 
This method simultaneously estimates the desirable HS image $\u$ and the sparse noise $\s$ for noise-robust restoration.
The first and second constraints measure data fidelities to the observation $\v$ and the sparse noise $\s$, respectively.
As mentioned in \cite{CSALSA, EPIpre, TSP2015Involved, WSNLRMA, LRTV, ASSTV_HSI2, TIP2017, HSSTV, ICASSP2018, APSIPA2018, ASSTV, ICASSP2019}, such a constraint-type data-fidelity enables us to translate the hyperparameter(s) balancing between regularization and data-fidelity like $\lambda_1$ and $\lambda_2$ in \eqref{prob:TV_based} to the upper bound of the degree of data-fidelity $\varepsilon$ and $\eta$ that can be set in a much easier manner. 

Since all constraints are closed convex sets, and HSSTV is a convex function, Prob.~\eqref{prob:HSSTV_Gaussian} is a constrained convex optimization problem.
In this paper, we adopt ADMM (see Sec.~\ref{subsec:ADMM}) for solving the problem.
In what follows, we reformulate Prob.~\eqref{prob:HSSTV_Gaussian} into Prob.~\eqref{ADMMequation}. 

By using the indicator functions of the constraints, Prob.~\eqref{prob:HSSTV_Gaussian} can be rewritten as
\begin{equation}\label{prob:HSSTVrewritten}
\min_{\u,\s} \|\A_{\omega}\u\|_{1,p} + \iota_{\Bmath^{\v}_{2,\varepsilon}}(\PHI\u + \s) + \iota_{\Bmath_{1,\eta}}(\s) + \iota_{[\mu_{\min}, \mu_{\max}]^{NB}}(\u).
\end{equation}
Note that from the definition of the indicator function, Prob.~\eqref{prob:HSSTVrewritten} exactly equals to Prob.~\eqref{prob:HSSTV_Gaussian}.
By letting
\begin{align}
&f:\R^{NB}\rightarrow\R^2:\u\mapsto (0, 0), \label{f} \\
&g:\R^{5NB+2M}\rightarrow\R\cup\{\infty\}: (\z_1,\z_2,\z_3,\z_4)\mapsto \nonumber \\
&\|\z_1\|_{1,p} + \iota_{\Bmath^{\v}_{2,\varepsilon}}(\z_2) + \iota_{\Bmath_{1,\varepsilon}}(\z_3) + \iota_{[\mu_{\min}, \mu_{\max}]^{NB}}(\z_4), \label{g} \\
&\G:\R^{NB}\rightarrow\R^{5NB+2M}:\u\mapsto(\A_{\omega}\u, \PHI\u + \s, \s, \u). \label{G}
\end{align}
Prob.~\eqref{prob:HSSTVrewritten} is reduced to Prob.~\eqref{ADMMequation}.
The resulting algorithm based on ADMM is summarized in Alg.~\ref{alg:ADMMHSSTV}.

\begin{algorithm}[t]
\footnotesize{
\LinesNumbered
\SetKwInOut{Input}{input}
\SetKwInOut{Output}{output}
\caption{ADMM method for Prob.~\eqref{prob:HSSTV_Gaussian}}
\label{alg:ADMMHSSTV}
\Input{$\z_1^{(0)}$, $\z_2^{(0)}$, $\z_3^{(0)}$, $\z_4^{(0)}$, $\d_1^{(0)}$, $\d_2^{(0)}$, $\d_3^{(0)}$, $\d_4^{(0)}$}
\vspace{3pt}
\While{A stopping criterion is not satisfied}{
\vspace{3pt} 
$(\u^{(n+1)}, \s^{(n+1)}) = \argmin_{\u,\s}\frac{1}{2\gamma} (\| \z_1^{(n)} - \A_{\omega}\u -\d_1^{(n)}\|^2 + \|\z_2^{(n)} - (\PHI\u + \s) - \d_2^{(n)}\|^2 + \|\z_3^{(n)} - \s -\d_3^{(n)}\|^2 + \|\z_4^{(n)} - \u - \d_4^{(n)}\|^2)$\;
$\z_1^{(n+1)} = \prox_{\gamma \|\cdot\|_{1,p}}(\A_{\omega}\u^{(n+1)} + \d_1^{(n)})$\;
$\z_2^{(n+1)} = \prox_{\gamma \iota_{\mathcal{B}^{\v}_{2,\varepsilon}}}(\PHI\u^{(n+1)} + \s^{(n+1)} + \d_2^{(n)})$\;
$\z_3^{(n+1)} = \prox_{\gamma \iota_{\mathcal{B}_{1,\eta}}}(\s^{(n+1)} + \d_3^{(n)})$\;
$\z_4^{(n+1)} = \prox_{\gamma \iota_{[\mu_{\min},\mu_{\max}]^{NB}}}(\u^{(n+1)} + \d_4^{(n)})$\;
$\d_1^{(n+1)} = \d_1^{(n)} + \A_{\omega}\u^{(n+1)} - \z_1^{(n+1)}$\;
$\d_2^{(n+1)} = \d_2^{(n)} + \PHI\u^{(n+1)} + \s^{(n+1)} - \z_2^{(n+1)}$\;
$\d_3^{(n+1)} = \d_3^{(n)} + \s^{(n+1)} - \z_3^{(n+1)}$\;
$\d_4^{(n+1)} = \d_4^{(n)} + \u^{(n+1)} - \z_4^{(n+1)}$\;
$n\leftarrow n+1$\;
}
}
\end{algorithm}

The update of $\u$ and $\s$ in Alg.~\ref{alg:ADMMHSSTV} come down to the following forms:
\begin{align}
\u^{(n+1)} =& \left(\A_{\omega}^{\top}\A_{\omega} + \PHI^{\top}\PHI + \frac{1}{2} \I \right)^{-1} \mbox{RHS}, \nonumber \\
\mbox{RHS} =& \A_{\omega}^{\top}(\z_1^{(n)} - \d_1^{(n)}) + \frac{1}{2}\PHI^{\top}(\z_2^{(n)} - \d_2^{(n)}) \nonumber \\
& - \frac{1}{2}(\z_3^{(n)} - \d_3^{(n)}) + (\z_4^{(n)} - \d_4^{(n)}), \label{update_u} \\
\s^{(n+1)} =& ~\frac{1}{2} (\z_2^{(n)} - \u^{(n+1)} - \d_2^{(n)} + \z_3^{(n)} - \d_3^{(n)}), \nonumber
\end{align}
Since the update of $\u$ and $\s$ in Alg.~\ref{alg:ADMMHSSTV} is strictly-convex quadratic minimization, one can obtain this update forms by differentiating it. 
Here, we should consider the structure of $\PHI$ because it affects the matrix inversion in \eqref{update_u}.
If $\PHI$ is a block-circulant-with-circulant-blocks (BCCB) matrix \cite{Hansen}, we can leverage 3DFFT to efficiently solve the inversion in Step 2 with the difference operators having periodic boundary,
i.e., $\A_{\omega}^{\top}\A_{\omega} + \PHI^{\top}\PHI + \I$ can be diagonalized by the 3D FFT matrix and its inverse.
If $\PHI$ is a semi-orthogonal matrix, i.e., $\PHI\PHI^{\top} = \alpha\I$ $(\alpha > 0)$,
we leave it to the update of $\z_2$,
which means that we replace $\iota_{\Bmath^{\v}_{2,\varepsilon}}$
by $\iota_{\Bmath^{\v}_{2,\varepsilon}} \circ \PHI$ in \eqref{g} and $\PHI\u$ by $\u$ in \eqref{G}.
This is because the proximity operator of $\iota_{\Bmath^{\v}_{2,\varepsilon}} \circ \PHI$ in this case can be computed by using \cite[Table 1.1-x]{TechCombettes} as follows:
\begin{equation*} 
\prox_{\gamma \iota_{\Bmath_{2,\varepsilon}^{\v}} \circ \PHI} (\x) = \x + \alpha^{-1}\PHI^{\top}(P_{\Bmath_{2,\varepsilon}^{\v}}(\PHI\x)-\PHI\x).
\end{equation*}
If $\PHI$ is a sparse matrix,
we offer to use a preconditioned conjugate gradient method \cite{PCG} for approximately solving the inversion,
or to apply primal-dual splitting methods \cite{PDChambolle,PDCombettes,PDCondat} instead of ADMM.\footnote{Primal-dual splitting methods require no matrix inversion but in general their convergence speed is slower than ADMM.}
Otherwise, randomized image restoration methods using stochastic proximal splitting algorithms \cite{ICASSP2016SDCAADMM, SPDSChambolle, SPDSCombettesOnlineIR, ICASSP2019Ono}
might be useful for reducing the computational cost.

For the update of $\z_1$, the proximity operators are reduced to simple soft-thresholding type operations:
for $\gamma > 0$ and for $i=1,\ldots,4NB$,
(i) in the case of $p = 1$, 
\begin{equation}
[\prox_{\gamma \|\cdot\|_1}(\x)]_i
= \sgn(x_i)\max \left \{ |x_i|-\gamma, 0 \right \}, \nonumber
\end{equation}
where $\sgn$ is the sign function, and (ii) in the case of $p = 2$, 
\begin{equation}
[\prox_{\gamma \|\cdot\|_{1,2}}(\x)]_i
= \max \left \{ 1 - \gamma \left( \sum_{j=0}^{3} x_{\tilde{i}+jNB}^2 \right )^{-\frac{1}{2}}, 0 \right \}
x_i, \nonumber
\end{equation}
where $\tilde{i} := ((i-1) \mod NB) + 1$.

The update of $\z_2$, $\z_3$, and $\z_4$ require the proximity operators of the indicator functions of $\Bmath_{2,\varepsilon}^{\v}$, $\Bmath_{1,\eta}$ and $[\mu_{\min}, \mu_{\max}]^{NB}$, respectively, which equal to the metric projections onto them (see Sec~\ref{subsec:prox}).
Specifically, the metric projection onto $\Bmath_{2,\varepsilon}^{\v}$ is given by
\begin{equation*}
P_{\Bmath_{2,\varepsilon}^{\v}}(\x) = \left \{
\begin{array}{l l}
\x, & \mbox{if } \x \in \Bmath_{2,\varepsilon}^{\v},\\
\v + \frac{\varepsilon(\x-\v)}{\|\x-\v\|}, & \mbox{otherwise},
\end{array} \right.
\end{equation*}
that onto $\Bmath_{1,\eta}$ is given by
\begin{equation*}
P_{\Bmath_{1,\eta}} = \sgn(\x)\max(|\x|-\eta, 0),
\end{equation*}
and that onto $[\mu_{\min}, \mu_{\max}]^{NB}$ is given, for $i = 1, \ldots, NB$, by
\begin{equation*}
[P_{[\mu_{\min}, \mu_{\max}]^{NB}}(\x)]_{i} = \min \{ \max \{ x_i, \mu_{\min} \} , \mu_{\max} \}.
\end{equation*}

\section{Experiments}\label{sec:E}

\begin{table}[t]
\begin{center}
\caption{Parameter settings for ASSTV, LRMR, LRTV, and the proposed method.} 
\label{table:settings}
\scalebox{0.88}{
\begin{tabular}{|c|c||c|c|}\hline
\multicolumn{2}{|c||}{\diagbox{ parameters }{noise level}} & (i) $(\sigma,~s_p,~l_v = l_h) = (0.05,~0.04,~0.04)$ & (ii) $(0.1,~0.05,~0.05)$ \\ \hline 
\multirow{2}{*}{ASSTV}& $\tau_v = \tau_h$ & \multicolumn{2}{c|}{1} \\ \cline{2-4}
& $\tau_b$ & 3 & 2 \\ \hline
\multirow{2}{*}{LRMR} & $r$ & \multicolumn{2}{c|}{3} \\ \cline{2-4}
& $k$ & \multicolumn{2}{c|}{$s_p + l_v + l_h - l_v l_h$ (the rate of sparse noise)} \\ \hline
\multirow{2}{*}{LRTV} & $r$ & \multicolumn{2}{c|}{2} \\ \cline{2-4}
& $\tau$ & 0.005 & 0.008 \\ \hline
{\bf{proposed}} & $\omega$ & \multicolumn{2}{c|}{0.04} \\ \hline
\end{tabular}}
\end{center}
\end{table}

\begin{table*}[tp]
\begin{center}
 \caption{PSNR (top) and SSIM (bottom) in mixed noise removal experiments.}
 \label{Mixed_data}
 \scalebox{0.86}{
 \begin{tabular}{|c||c|c||c|c|c|c|c|c|m{13mm}|m{13mm}|} \hline
& HS image & noise level & HTV & SSAHTV & SSTV & ASSTV & LRMR & LRTV & {\bf{proposed}} {\bf{($p = 1$)}} & {\bf{proposed}} {\bf{($p = 2$)}} \\ \cline{1-11}
& Beltsville & (i) & 29.43 & 29.47 & 33.66 & 27.16 & 30.91 & {\bf{35.32}} & \multicolumn{1}{c|}{34.25} & \multicolumn{1}{c|}{34.16} \\ 
 & $256 \times 256 \times 32$ & (ii) & 26.40 & 26.43 & 28.42 & 24.60 & 27.13 & {\bf{31.22}} & \multicolumn{1}{c|}{29.79} & \multicolumn{1}{c|}{29.62} \\ \cline{2-11}
 & Suwannee & (i) & 30.14 & 30.18 & 34.59 & 32.60 & 30.30 & {\bf{36.20}} & \multicolumn{1}{c|}{35.15} & \multicolumn{1}{c|}{36.01}\\ 
  & $256 \times 256 \times 32$ & (ii) & 26.70 & 26.74 & 29.55 & 28.71 & 26.90 & {\bf{31.95}} & \multicolumn{1}{c|}{31.08} & \multicolumn{1}{c|}{31.22} \\ \cline{2-11}
 & DC & (i) & 26.46 & 26.51 & 33.03 & 28.80 & 31.71 & {\bf{34.78}} & \multicolumn{1}{c|}{33.36} & \multicolumn{1}{c|}{33.08} \\ 
 & $256 \times 256 \times 32$ & (ii) & 23.84 & 23.88 & 27.71 & 25.25 & 27.35 & {\bf{29.53}} & \multicolumn{1}{c|}{28.57} & \multicolumn{1}{c|}{28.32} \\ \cline{2-11}
 & Cuprite & (i) & 31.67 & 31.68 & 34.42 & 29.14 & 30.16 & 28.39 & \multicolumn{1}{c|}{34.96} & \multicolumn{1}{c|}{{\bf{36.20}}} \\ 
 & $256 \times 256 \times 32$ & (ii) & 28.20 & 28.21 & 29.86 & 26.57 & 27.32 & 27.94 & \multicolumn{1}{c|}{31.63} & \multicolumn{1}{c|}{{\bf{31.73}}} \\ \cline{2-11}
 & Reno & (i) & 28.53 & 28.57 & 34.37 & 30.49 & 32.21 & {\bf{37.06}} & \multicolumn{1}{c|}{35.11} & \multicolumn{1}{c|}{34.96} \\ 
& $256 \times 256 \times 32$ & (ii) & 25.56 & 25.61 & 28.11 & 26.95 & 28.47 & {\bf{31.00}} & \multicolumn{1}{c|}{29.83} & \multicolumn{1}{c|}{29.72} \\ \cline{2-11}
 & Botswana & (i) & 27.98 & 28.05 & 33.32 & 26.47 & 31.62 & 29.00 & \multicolumn{1}{c|}{{\bf{33.61}}} & \multicolumn{1}{c|}{33.53} \\ 
  & $256 \times 256 \times 32$ & (ii) & 25.21 & 25.25 & 28.55 & 24.01 & 28.31 & 27.33 & \multicolumn{1}{c|}{{\bf{29.39}}} & \multicolumn{1}{c|}{29.35} \\ \cline{2-11}
 PSNR & IndianPines & (i) & 31.05 & 31.06 & 31.45 & 29.07 & 28.96 & 26.16 & \multicolumn{1}{c|}{{\bf{31.90}}} & \multicolumn{1}{c|}{31.80} \\ 
 & $145 \times 145 \times 32$ & (ii) & 28.57 & 28.57 & 27.82 & 26.72 & 25.14 & {\bf{29.82}} & \multicolumn{1}{c|}{29.26} & \multicolumn{1}{c|}{29.18} \\ \cline{2-11}
 & KSC & (i) & 30.17 & 30.25 & 34.74 & 31.64 & 33.74 & 35.74 & \multicolumn{1}{c|}{{\bf{36.39}}} & \multicolumn{1}{c|}{36.33} \\ 
 & $256 \times 256 \times 32$ & (ii) & 28.03 & 28.06 & 29.23 & 28.62 & 30.19 & 30.22 & \multicolumn{1}{c|}{{\bf{31.82}}} & \multicolumn{1}{c|}{31.72} \\ \cline{2-11}
 & PaviaLeft & (i) & 27.62 & 27.70 & 35.57 & 30.91 & 33.01 & 36.49 & \multicolumn{1}{c|}{{\bf{35.98}}} & \multicolumn{1}{c|}{35.81} \\ 
 & $216 \times 216 \times 32$ & (ii) & 24.74 & 24.78 & 29.93 & 26.71 & 29.46 & 29.02 & \multicolumn{1}{c|}{{\bf{30.47}}} & \multicolumn{1}{c|}{30.24} \\ \cline{2-11}
 & PaviaRight & (i) & 26.93 & 27.35 & 34.54 & 31.13 & 33.33 & {\bf{35.82}} & \multicolumn{1}{c|}{35.68} & \multicolumn{1}{c|}{35.23} \\ 
 & $256 \times 256 \times 32$ & (ii) & 24.90 & 25.16 & 30.70 & 27.23 & 29.82 & 29.08 & \multicolumn{1}{c|}{{\bf{31.59}}} & \multicolumn{1}{c|}{31.39} \\ \cline{2-11}
 & PaviaU & (i) & 27.92 & 28.04 & 35.52 & 31.65 & 33.00 & {\bf{36.72}} & \multicolumn{1}{c|}{36.31} & \multicolumn{1}{c|}{36.17} \\ 
 & $256 \times 256 \times 32$ & (ii) & 25.24 & 25.29 & 30.21 & 27.42 & 29.43 & 28.90 & \multicolumn{1}{c|}{{\bf{31.04}}} & \multicolumn{1}{c|}{30.80} \\ \cline{2-11}
 & Salinas & (i) & 32.59 & 32.64 & 35.86 & 32.83 & 31.82 & 36.74 & \multicolumn{1}{c|}{37.60} & \multicolumn{1}{c|}{{\bf{37.65}}} \\ 
  & $217 \times 217 \times 32$ & (ii) & 28.88 & 28.91 & 28.19 & 28.99 & 28.02 & {\bf{32.73}} & \multicolumn{1}{c|}{32.01} & \multicolumn{1}{c|}{32.12} \\ \cline{2-11}
  & SalinaA & (i) & 32.54 & 32.65 & 35.29 & 28.12 & 31.18 & 28.49 & \multicolumn{1}{c|}{{\bf{36.27}}} & \multicolumn{1}{c|}{36.23} \\
 & $83 \times 86 \times 32$ & (ii) & 28.69 & 28.80 & 29.67 & 25.19 & 27.67 & 26.10 & \multicolumn{1}{c|}{{\bf{31.68}}} & \multicolumn{1}{c|}{31.64} \\ \hline \hline
 & & (i) & 0.7902 & 0.7904 & 0.8856 & 0.8111 & 0.8583 & {\bf{0.9372}} & \multicolumn{1}{c|}{0.9132} & \multicolumn{1}{c|}{0.9085} \\
 & Beltsville & (ii) & 0.6954 & 0.6959 & 0.7057 & 0.7177 & 0.7083 & {\bf{0.8568}} & \multicolumn{1}{c|}{0.8186} & \multicolumn{1}{c|}{0.8088} \\ \cline{2-11}
  & & (i) & 0.8406 & 0.8410 & 0.9353 & 0.9052 & 0.8689 & 0.9502 & \multicolumn{1}{c|}{{\bf{0.9559}}} & \multicolumn{1}{c|}{0.9555} \\ 
  & Suwannee & (ii) & 0.7542 & 0.7552 & 0.8146 & 0.8226 & 0.7470 & 0.8930 & \multicolumn{1}{c|}{0.9125} & \multicolumn{1}{c|}{{\bf{0.9158}}} \\ \cline{2-11}
  & & (i) & 0.7622 & 0.7633 & 0.9274 & 0.8676 & 0.9248 & {\bf{0.9613}} & \multicolumn{1}{c|}{0.9442} & \multicolumn{1}{c|}{0.9394} \\ 
 & DC & (ii) & 0.6189 & 0.6201 & 0.8092 & 0.7211 & 0.8214 & {\bf{0.8810}} & \multicolumn{1}{c|}{0.8611} & \multicolumn{1}{c|}{0.8533} \\ \cline{2-11}
 & & (i) & 0.8550 & 0.8552 & 0.9179 & 0.8632 & 0.8495 & 0.9396 & \multicolumn{1}{c|}{{\bf{0.9459}}} & \multicolumn{1}{c|}{0.9426} \\ 
 & Cuprite & (ii) & 0.7849 & 0.7852 & 0.7717 & 0.7953 & 0.7098 & 0.8814 & \multicolumn{1}{c|}{0.9031} & \multicolumn{1}{c|}{{\bf{0.9058}}} \\ \cline{2-11}
 & & (i) & 0.7818 & 0.7819 & 0.9322 & 0.8832 & 0.9012 & {\bf{0.9589}} & \multicolumn{1}{c|}{0.9531} & \multicolumn{1}{c|}{0.9515} \\ 
 & Reno & (ii) & 0.6640 & 0.6645 & 0.8045 & 0.7539 & 0.7905 & {\bf{0.8816}} & \multicolumn{1}{c|}{0.8679} & \multicolumn{1}{c|}{0.8635} \\ \cline{2-11}
 & & (i) & 0.7896 & 0.7900 & 0.9202 & 0.8199 & 0.9068 & 0.9282 & \multicolumn{1}{c|}{0.9343} & \multicolumn{1}{c|}{{\bf{0.9344}}} \\ 
 & Botswana & (ii) & 0.6810 & 0.6820 & 0.8175 & 0.7095 & 0.8201 & 0.8564 & \multicolumn{1}{c|}{0.8745} & \multicolumn{1}{c|}{{\bf{0.8765}}} \\ \cline{2-11}
 & & (i) & 0.8118 & 0.8120 & 0.8015 & 0.7671 & 0.7593 & 0.8190 & \multicolumn{1}{c|}{0.8335} & \multicolumn{1}{c|}{0.8243} \\ 
 SSIM & IndianPines & (ii) & 0.7713 & 0.7713 & 0.6229 & 0.7303 & 0.7893 & {\bf{0.7939}} & \multicolumn{1}{c|}{0.7785} & \multicolumn{1}{c|}{0.7689} \\ \cline{2-11}
& & (i) & 0.8271 & 0.8278 & 0.9116 & 0.8922 & 0.8890 & 0.9385 & \multicolumn{1}{c|}{{\bf{0.9542}}} & \multicolumn{1}{c|}{0.9532} \\ 
 & KSC & (ii) & 0.7598 & 0.7602 & 0.7885 & 0.8064 & 0.7529 & 0.8427 & \multicolumn{1}{c|}{{\bf{0.8809}}} & \multicolumn{1}{c|}{0.8747} \\ \cline{2-11}
 & & (i) & 0.7752 & 0.7770 & 0.9593 & 0.8828 & 0.9359 & 0.9612 & \multicolumn{1}{c|}{{\bf{0.9661}}} & \multicolumn{1}{c|}{0.9645} \\ 
 & PaviaLeft & (ii) & 0.6102 & 0.6116 & 0.8755 & 0.7267 & 0.8565 & 0.8791 & \multicolumn{1}{c|}{{\bf{0.8898}}} & \multicolumn{1}{c|}{0.8815} \\ \cline{2-11}
 & & (i) & 0.7769 & 0.7772 & 0.9494 & 0.8862 & 0.9256 & 0.9540 & \multicolumn{1}{c|}{{\bf{0.9616}}} & \multicolumn{1}{c|}{0.9598} \\ 
 & PaviaRight & (ii) & 0.6474 & 0.6471 & 0.8635 & 0.7493 & 0.8261 & 0.8507 & \multicolumn{1}{c|}{{\bf{0.9086}}} & \multicolumn{1}{c|}{0.9006} \\ \cline{2-11}
 & & (i) & 0.7973 & 0.7986 & 0.9452 & 0.8891 & 0.9124 & 0.9540 & \multicolumn{1}{c|}{{\bf{0.9622}}} & \multicolumn{1}{c|}{0.9610} \\ 
 & PaviaU & (ii) & 0.6776 & 0.6785 & 0.8444 & 0.7678 & 0.8103 & 0.8627 & \multicolumn{1}{c|}{{\bf{0.8935}}} & \multicolumn{1}{c|}{0.8855} \\ \cline{2-11}
 & & (i) & 0.8997 & 0.9002 & 0.9015 & 0.9163 & 0.8270 & 0.9509 & \multicolumn{1}{c|}{0.9561} & \multicolumn{1}{c|}{{\bf{0.9564}}} \\ 
 & Salinas & (ii) & 0.8570 & 0.8575 & 0.7117 & 0.8732 & 0.6670 & 0.9225 & \multicolumn{1}{c|}{0.9223} & \multicolumn{1}{c|}{{\bf{0.9240}}} \\ \cline{2-11}
  & & (i) & 0.9129 & 0.9137 & 0.9134 & 0.8468 & 0.8632 & 0.9384 & \multicolumn{1}{c|}{{\bf{0.9448}}} & \multicolumn{1}{c|}{0.9416} \\
 & SalinaA & (ii) & 0.8793 & 0.8803 & 0.7789 & 0.8110 & 0.7266 & 0.8951 & \multicolumn{1}{c|}{{\bf{0.9197}}} & \multicolumn{1}{c|}{0.9195} \\ \cline{1-11}
 \end{tabular}}
\end{center}
\end{table*}

\begin{figure}[t]
\begin{center}
\begin{minipage}[t]{0.16\hsize}
 \includegraphics[bb = 0 0 217 217, width=\hsize]{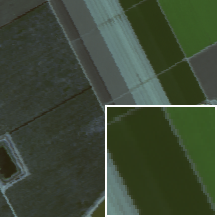}
\end{minipage}
\begin{minipage}[t]{0.16\hsize}
 \includegraphics[bb = 0 0 217 217, width=\hsize]{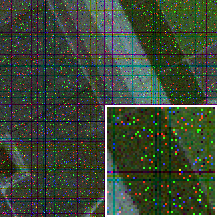}
\end{minipage}
 \begin{minipage}[t]{0.16\hsize}
 \includegraphics[bb = 0 0 217 217, width=\hsize]{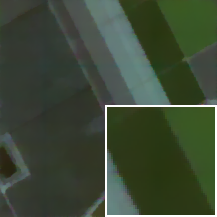}
\end{minipage}
 \begin{minipage}[t]{0.16\hsize}
 \includegraphics[bb = 0 0 217 217, width=\hsize]{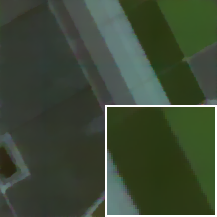}
\end{minipage}
 \begin{minipage}[t]{0.16\hsize}
 \includegraphics[bb = 0 0 217 217, width=\hsize]{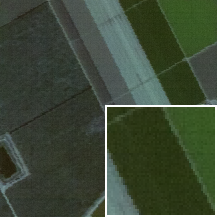}
\end{minipage}
 \begin{minipage}[t]{0.16\hsize}
 \includegraphics[bb = 0 0 217 217, width=\hsize]{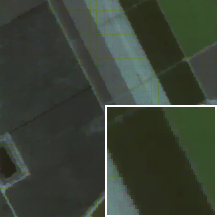}
\end{minipage}

\vspace{-3pt}
 \begin{minipage}[t]{0.16\hsize}
\centerline{\footnotesize{ Salinas }}
\end{minipage}
 \begin{minipage}[t]{0.16\hsize}
\centerline{\footnotesize{ 17.06, 0.1908 }}
\end{minipage}
 \begin{minipage}[t]{0.16\hsize}
\centerline{\footnotesize{ 32.59, 0.8997 }}
\end{minipage}
\begin{minipage}[t]{0.16\hsize}
\centerline{\footnotesize{ 32.66, 0.9003 }}
\end{minipage}
 \begin{minipage}[t]{0.16\hsize}
\centerline{\footnotesize{ 35.86, 0.9015 }}
\end{minipage}
 \begin{minipage}[t]{0.16\hsize}
\centerline{\footnotesize{ 32.83, 0.9163 }}
\end{minipage}

\vspace{-3pt}
 \begin{minipage}[t]{0.16\hsize}
\centerline{\footnotesize{ groundtruth }}
\end{minipage}
 \begin{minipage}[t]{0.16\hsize}
\centerline{\footnotesize{ observation }}
\end{minipage}
 \begin{minipage}[t]{0.16\hsize}
\centerline{\footnotesize{ HTV }}
\end{minipage}
 \begin{minipage}[t]{0.16\hsize}
\centerline{\footnotesize{ SSAHTV }}
\end{minipage}
 \begin{minipage}[t]{0.16\hsize}
\centerline{\footnotesize{ SSTV }}
\end{minipage}
 \begin{minipage}[t]{0.16\hsize}
\centerline{\footnotesize{ ASSTV }}
\end{minipage}

\vspace{3pt}
 \begin{minipage}[t]{0.16\hsize}
 \centerline{\footnotesize{ }}
\end{minipage}
 \begin{minipage}[t]{0.16\hsize}
 \centerline{\footnotesize{ }}
\end{minipage}
 \begin{minipage}[t]{0.16\hsize}
 \includegraphics[bb = 0 0 217 217, width=\hsize]{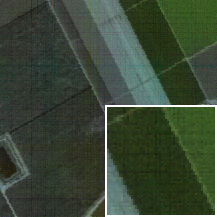}
\end{minipage}
 \begin{minipage}[t]{0.16\hsize}
 \includegraphics[bb = 0 0 217 217, width=\hsize]{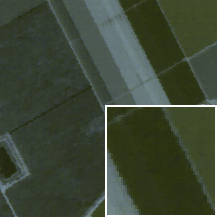}
\end{minipage}
 \begin{minipage}[t]{0.16\hsize}
 \includegraphics[bb = 0 0 217 217, width=\hsize]{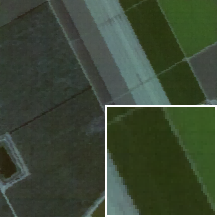}
\end{minipage}
 \begin{minipage}[t]{0.16\hsize}
 \includegraphics[bb = 0 0 217 217, width=\hsize]{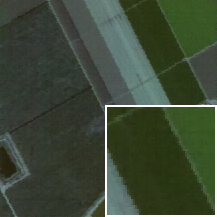}
\end{minipage}

\vspace{-3pt}
\begin{minipage}[t]{0.16\hsize}
 \centerline{\footnotesize{ }}
\end{minipage}
\begin{minipage}[t]{0.16\hsize}
 \centerline{\footnotesize{ }}
\end{minipage}
 \begin{minipage}[t]{0.16\hsize}
 \centerline{\footnotesize{ 31.82, 0.8270 }}
\end{minipage}
 \begin{minipage}[t]{0.16\hsize}
\centerline{\footnotesize{ 36.74, 0.9509 }}
\end{minipage}
 \begin{minipage}[t]{0.16\hsize}
\centerline{\footnotesize{ 37.60, 0.9561 }}
\end{minipage}
 \begin{minipage}[t]{0.16\hsize}
\centerline{\footnotesize{ {\bf{37.65}}, {\bf{0.9564}} }}
\end{minipage}

\vspace{-3pt}
\begin{minipage}[t]{0.16\hsize}
 \centerline{\footnotesize{ }}
\end{minipage}
\begin{minipage}[t]{0.16\hsize}
 \centerline{\footnotesize{ }}
\end{minipage}
 \begin{minipage}[t]{0.16\hsize}
 \centerline{\footnotesize{ LRMR }}
\end{minipage}
 \begin{minipage}[t]{0.16\hsize}
\centerline{\footnotesize{ LRTV }}
\end{minipage}
\begin{minipage}[t]{0.16\hsize}
\centerline{\footnotesize{ {\bf{proposed}} }}
\centerline{\footnotesize{ {\bf{($p = 1$)}} }}
\end{minipage}
 \begin{minipage}[t]{0.16\hsize}
\centerline{\footnotesize{ {\bf{proposed}} }}
\centerline{\footnotesize{ {\bf{($p = 2$)}} }}
\end{minipage}

\vspace{5pt}
 \begin{minipage}[t]{0.16\hsize}
 \includegraphics[bb = 0 0 256 256, width=\hsize]{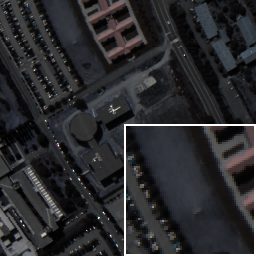}
\end{minipage}
\begin{minipage}[t]{0.16\hsize}
 \includegraphics[bb = 0 0 256 256, width=\hsize]{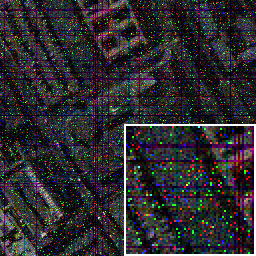}
\end{minipage}
 \begin{minipage}[t]{0.16\hsize}
 \includegraphics[bb = 0 0 256 256, width=\hsize]{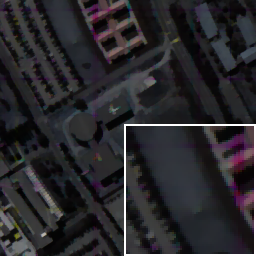}
\end{minipage}
 \begin{minipage}[t]{0.16\hsize}
 \includegraphics[bb = 0 0 256 256, width=\hsize]{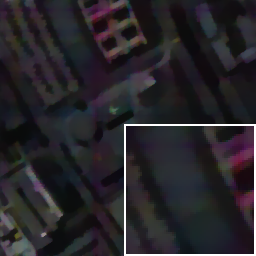}
\end{minipage}
 \begin{minipage}[t]{0.16\hsize}
 \includegraphics[bb = 0 0 256 256, width=\hsize]{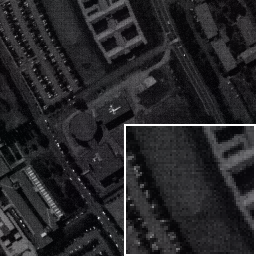}
\end{minipage}
 \begin{minipage}[t]{0.16\hsize}
 \includegraphics[bb = 0 0 256 256, width=\hsize]{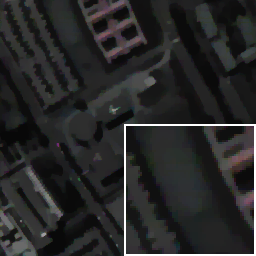}
\end{minipage}

\vspace{-3pt}
 \begin{minipage}[t]{0.16\hsize}
\centerline{\footnotesize{ PaviaU }}
\end{minipage}
 \begin{minipage}[t]{0.16\hsize}
\centerline{\footnotesize{ 15.39, 0.1877 }}
\end{minipage}
 \begin{minipage}[t]{0.16\hsize}
\centerline{\footnotesize{ 25.24, 0.6776 }}
\end{minipage}
 \begin{minipage}[t]{0.16\hsize}
\centerline{\footnotesize{ 25.29, 0.6785 }}
\end{minipage}
 \begin{minipage}[t]{0.16\hsize}
\centerline{\footnotesize{ 30.21, 0.8444 }}
\end{minipage}
 \begin{minipage}[t]{0.16\hsize}
\centerline{\footnotesize{ 27.42, 0.7678 }}
\end{minipage}

\vspace{-3pt}
 \begin{minipage}[t]{0.16\hsize}
\centerline{\footnotesize{ groundtruth }}
\end{minipage}
 \begin{minipage}[t]{0.16\hsize}
\centerline{\footnotesize{ observation }}
\end{minipage}
 \begin{minipage}[t]{0.16\hsize}
\centerline{\footnotesize{ HTV }}
\end{minipage}
 \begin{minipage}[t]{0.16\hsize}
\centerline{\footnotesize{ SSAHTV }}
\end{minipage}
 \begin{minipage}[t]{0.16\hsize}
\centerline{\footnotesize{ SSTV }}
\end{minipage}
 \begin{minipage}[t]{0.16\hsize}
\centerline{\footnotesize{ ASSTV }}
\end{minipage}

\vspace{3pt}
\begin{minipage}[t]{0.16\hsize}
 \centerline{\footnotesize{ }}
\end{minipage}
\begin{minipage}[t]{0.16\hsize}
 \centerline{\footnotesize{ }}
\end{minipage}
 \begin{minipage}[t]{0.16\hsize}
 \includegraphics[bb = 0 0 256 256, width=\hsize]{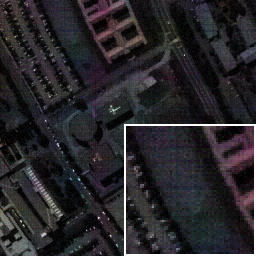}
\end{minipage}
 \begin{minipage}[t]{0.16\hsize}
 \includegraphics[bb = 0 0 256 256, width=\hsize]{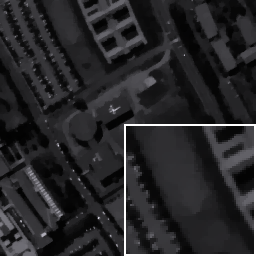}
\end{minipage}
 \begin{minipage}[t]{0.16\hsize}
 \includegraphics[bb = 0 0 256 256, width=\hsize]{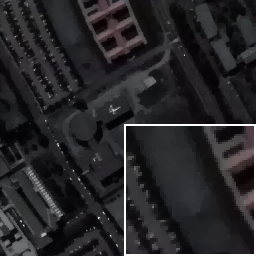}
\end{minipage}
 \begin{minipage}[t]{0.16\hsize}
 \includegraphics[bb = 0 0 256 256, width=\hsize]{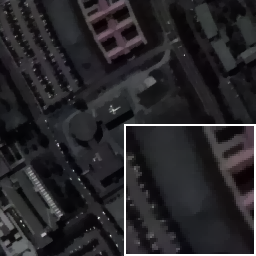}
\end{minipage}

\vspace{-3pt}
\begin{minipage}[t]{0.16\hsize}
 \centerline{\footnotesize{ }}
\end{minipage}
\begin{minipage}[t]{0.16\hsize}
 \centerline{\footnotesize{ }}
\end{minipage}
 \begin{minipage}[t]{0.16\hsize}
 \centerline{\footnotesize{ 29.43, 0.8103 }}
\end{minipage}
 \begin{minipage}[t]{0.16\hsize}
\centerline{\footnotesize{ 28.90, 0.8627 }}
\end{minipage}
\begin{minipage}[t]{0.16\hsize}
\centerline{\footnotesize{ {\bf{31.04}}, {\bf{0.8935}} }}
\end{minipage}
 \begin{minipage}[t]{0.16\hsize}
\centerline{\footnotesize{ 30.80, 0.8855 }}
\end{minipage}

\vspace{-3pt}
\begin{minipage}[t]{0.16\hsize}
 \centerline{\footnotesize{ }}
\end{minipage}
\begin{minipage}[t]{0.16\hsize}
 \centerline{\footnotesize{ }}
\end{minipage}
 \begin{minipage}[t]{0.16\hsize}
 \centerline{\footnotesize{ LRMR }}
\end{minipage}
 \begin{minipage}[t]{0.16\hsize}
\centerline{\footnotesize{ LRTV }}
\end{minipage}
\begin{minipage}[t]{0.16\hsize}
\centerline{\footnotesize{ {\bf{proposed}} }}
\centerline{\footnotesize{ {\bf{($p = 1$)}} }}
\end{minipage}
 \begin{minipage}[t]{0.16\hsize}
\centerline{\footnotesize{ {\bf{proposed}} }}
\centerline{\footnotesize{ {\bf{($p = 2$)}} }}
\end{minipage}

 \caption{Resulting HS images with their PSNR (left) and SSIM (right) in the mixed noise removal experiment (top: Salinas, the noise level (i), bottom: PaviaU, the noise level (ii)).}
 \label{fig:img_mixed}
 \vspace{-5pt}
\end{center}
\end{figure}

We demonstrate the advantages of the proposed method by applying it to two specific HS image restoration problems: denoising and CS reconstruction. 
In these experiments, we used 13 HS images taken from the \textit{SpecTIR} \cite{SpecTIR}, \textit{MultiSpec} \cite{MultiSpec} and \textit{GIC} \cite{GIC}, where their dynamic range were normalized into $[0,1]$.

The proposed method was compared with HTV~\cite{HTV}, SSAHTV~\cite{HTV}, SSTV~\cite{SSTV}, and ASSTV~\cite{ASSTV}. 
For a fair comparison, we replaced HSSTV in Prob.~\eqref{prob:HSSTV_Gaussian} with HTV, SSAHTV, SSTV, or ASSTV and solved the problem by ADMM.
In the denoising experiments, we also compared our proposed method with LRMR~\cite{LRMR} and TV-regularized low-rank matrix factorization (LRTV)~\cite{LRTV}.
Since LRMR and LRTV are customized to the mixed noise removal problem, we cannot adopt them for CS reconstruction.
We did not compare our proposed method with a recent CNN-based HS image denoising method \cite{3DADCNN}.
The CNN-based method cannot be represented as explicit regularization functions and is fully customized to denoising tasks.
In contrast, our proposed method can be used as a building block in various HS image restoration methods based on optimization.
Meanwhile, CNN-based methods strongly depend on what training data are used, which means that they cannot adapt to a wide range of noise intensity.
Thus, the design concepts of these methods are different from TVs and LRM-based approaches.

To quantitively evaluate restoration performance, we used the peak signal-to-noise ratio (PSNR) [dB] index and the structural similarity (SSIM) \cite{SSIM} index between a true HS image $\bar\u$ and a restored HS image $\u$.
PSNR is defined by $10\log_{10}(NB/\|\u-\bar\u\|^2)$, and the higher the value is, the more similar the two images are. 
SSIM is an image quality assessment index based on the human vision system, which is defined as follows:
\begin{align*}
\mbox{SSIM}(\u,\bar{\u}) &= \frac{1}{P}\sum_{i = 1}^{P} \mbox{SSIM}_i (\u,\bar{\u}), \\
\mbox{SSIM}_i (\u,\bar{\u}) &= \frac{(2\mu_{\u_i}\mu_{\bar{\u}_i} + C_1)(2\sigma_{\u_i\bar{\u}_i} + C_2)}{(\mu_{\u_i}^2 + \mu_{\bar{\u}_i}^2 + C_1)(\sigma_{\u_i}^2 + \sigma_{\bar{\u}_i}^2 + C_2)},
\end{align*}
where $\u_i$ and $\bar{\u}_i$ are the $i$th pixel-centered local patches of a restored HS image and a true HS image, respectively, $P$ is the number of patches, $\mu_{\u_i}$ and $\mu_{\bar{\u}_i}$ is the average values of the local patches of the restored and true HS images, respectively, 
$\sigma_{\u_i}$ and $\sigma_{\bar{\u}_i}$ represent the variances of $\u_i$ and $\bar{\u}_i$, respectively, and $\sigma_{\u_i \bar{\u}_i}$ denotes the covariance between $\u_i$ and $\bar{\u}_i$. 
Moreover, $C_1$ and $C_2$ are two constants, which avoid the numerical instability when either $\mu_{\u_i}^2 + \mu_{\bar{\u}_i}^2$ or $\sigma_{\u_i}^2 + \sigma_{\bar{\u}_i}^2$ is very close to zero.
SSIM gives a normalized score between zero and one, where the maximum value means that $\u$ equals to $\bar{\u}$.

We set the max iteration number, the stepsize $\gamma$ and the stopping criterion of ADMM to 10000, 0.05 and $\|\u^{(n)} - \u^{(n+1)}\| < 0.01$, respectively.

\begin{figure}[t]
\begin{center}
 \begin{minipage}[t]{0.24\hsize}
 \includegraphics[bb = 0 0 560 420, width=1.0\hsize]{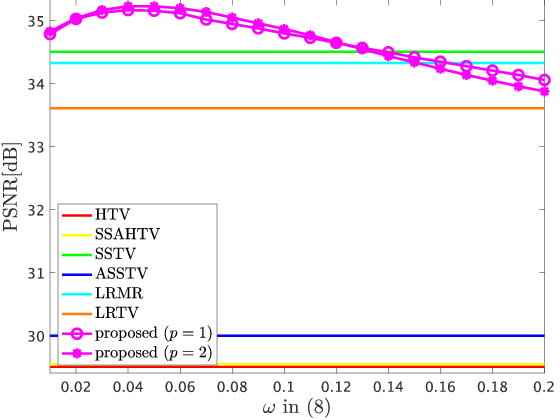}
 \end{minipage}
 \begin{minipage}[t]{0.24\hsize}
 \includegraphics[bb = 0 0 560 420, width=1.0\hsize]{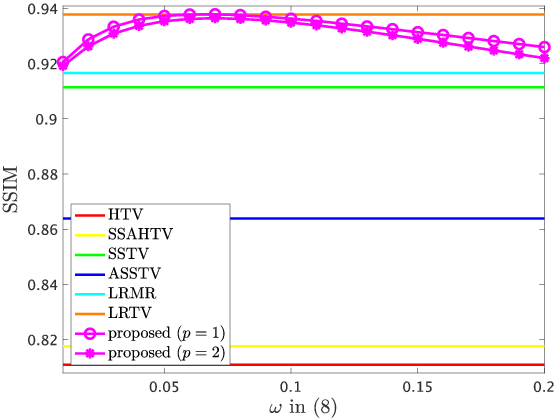}
 \end{minipage}
  \begin{minipage}[t]{0.24\hsize}
 \includegraphics[bb = 0 0 560 420, width=1.0\hsize]{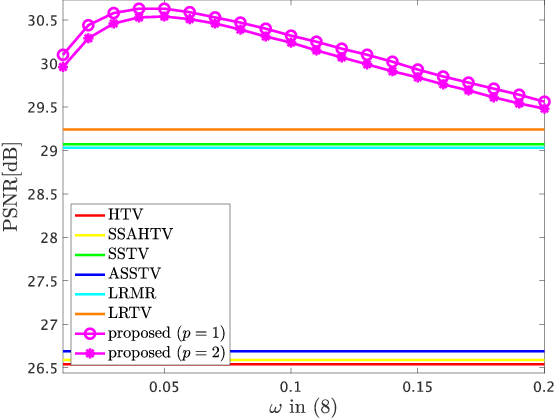}
 \end{minipage}
 \begin{minipage}[t]{0.24\hsize}
 \includegraphics[bb = 0 0 560 420, width=1.0\hsize]{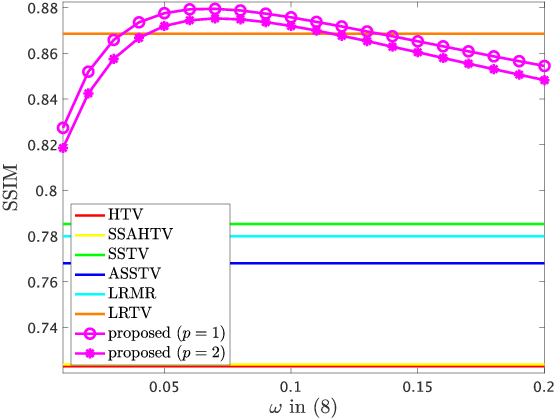}
 \end{minipage}
  
 \vspace{-3pt}
 \begin{minipage}[t]{0.24\hsize}
 \centerline{\footnotesize{PSNR, noise level (i)}}
 \end{minipage}
 \begin{minipage}[t]{0.24\hsize}
 \centerline{\footnotesize{SSIM, noise level (i)}}
 \end{minipage}
 \begin{minipage}[t]{0.24\hsize}
 \centerline{\footnotesize{PSNR, noise level (ii)}}
 \end{minipage}
 \begin{minipage}[t]{0.24\hsize}
 \centerline{\footnotesize{SSIM, noise level (ii)}}
 \end{minipage}
 
 \caption{PSNR or SSIM versus $\omega$ in \eqref{HSSTV} in the mixed noise removal experiment.}
 \label{omega_graph_mixed}
\end{center}
\end{figure}

\begin{figure}[t]
\begin{center}
 \begin{minipage}[t]{0.24\hsize}
 \includegraphics[bb = 0 0 560 420, width=1.0\hsize]{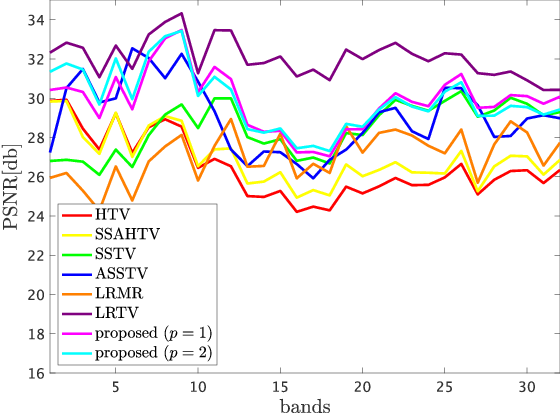}
 \centerline{\footnotesize{(a) Bandwise PSNR}}
 \end{minipage}
 \begin{minipage}[t]{0.24\hsize}
 \includegraphics[bb = 0 0 560 420, width=1.0\hsize]{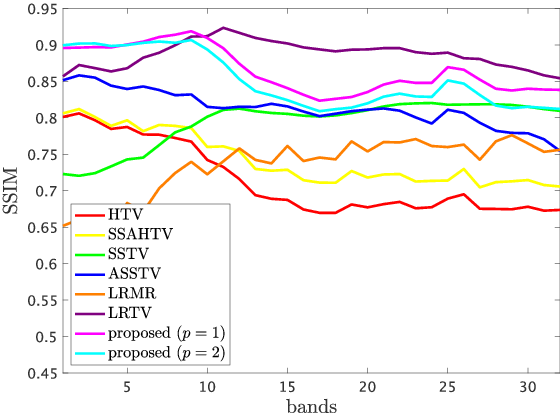}
 \centerline{\footnotesize{(b) Bandwise SSIM}}
 \end{minipage}
 \begin{minipage}[t]{0.24\hsize}
 \includegraphics[bb = 0 0 560 420, width=1.0\hsize]{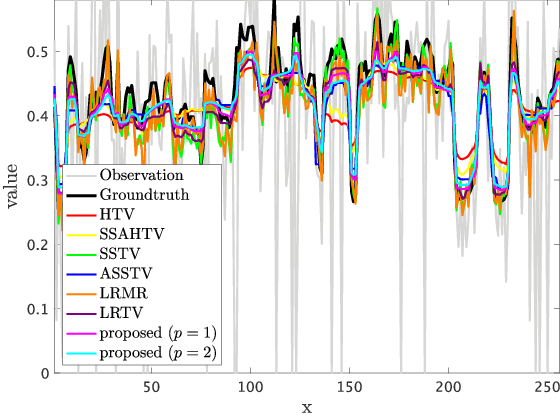}
 \centerline{\footnotesize{(c) Spatial response}}
 \end{minipage}
 \begin{minipage}[t]{0.24\hsize}
 \includegraphics[bb = 0 0 560 420, width=1.0\hsize]{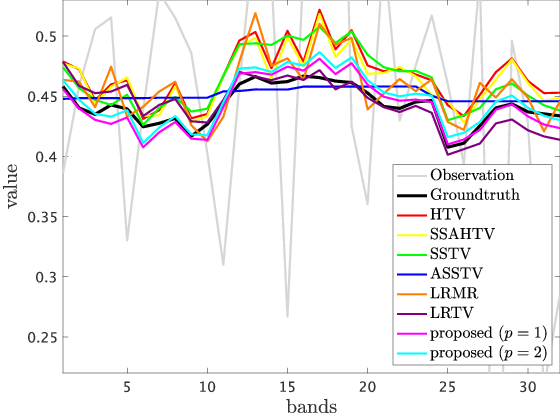}
 \centerline{\footnotesize{(d) Spectral response}}
 \end{minipage}
 \caption{ Bandwise PSNR and SSIM and spatial and spectral responses in the mixed noise removal experiment (Suwannee).}
 \label{other_graph_Gaussian}
\end{center}
\end{figure}

\subsection{Denoising}\label{subsec:ex_denoising}
First, we experimented on Gaussian-sparse mixed noise removal of HS images, where observed HS images included an additive white Gaussian noise $\n$ with the standard deviation $\sigma$ and sparse noise $\s$. 
In these experiments, we assumed that sparse noise consists of salt-and-pepper noise and vertical and horizontal line noise with these noise ratio in all pixels is $s_p$, $l_v$, and $l_h$, respectively.
We generated noisy HS images by adding two types of mixed noise to true HS images: (i) $(\sigma, ~s_p, ~l_v = l_h) = (0.05,~ 0.04,~ 0.04)$, (ii) $(0.1,~ 0.05,~ 0.05)$.
In the denoising case, $\PHI = \I$ in \eqref{model}, and the radiuses $\varepsilon$ and $\eta$ in Prob.~\eqref{prob:HSSTV_Gaussian} were set to $0.83 \sqrt{NB(1-(s_p(1-l_v-l_h) + l_v + l_h - l_v l_h))\sigma^2}$ and $ NB(0.45 s_p + (l_v + l_h) v_{ave} - l_v l_h v_{ave})$, respectively, where $v_{ave}$ is the average of the observed image.
Table~\ref{table:settings} shows the parameters settings for ASSTV, LRMR, LRTV, and the proposed method. 
We set these parameters to achieve the best performance for each method.

In Tab.~\ref{Mixed_data}, we show PSNR and SSIM of the denoised HS images by each method for two types of noise intensity and HS images.
For HTV, SSAHTV, SSTV, ASSTV, and LRMR, the proposed method outperforms the existing methods.
In the LRTV case, one can see that some results by LRTV outperform them by the proposed method. 
LRTV utilizes both TV-based and LRM-based regularization techniques, leading to higher-quality restoration than the proposed method.
Meanwhile, even though the proposed method uses only TV-based regularization, it outperforms LRTV over half situations.


Fig.~\ref{fig:img_mixed} shows the resulting images on \textit{Salinas} (the noise level (i), top) and \textit{PaviaU} (the noise level (ii), bottom) with their PSNR (left) and SSIM (right). 
Here, we depicted these HS images as RGB images (R = 8th, G = 16th, and B = 32nd bands). 
One can see that the results by HTV, SSAHTV, and ASSTV lose spacial details, and noise remains in the results by SSTV and LRMR.
Besides, since the restored images by SSTV and LRTV lose color with large noise intensity, SSTV and LRTV change spectral variation.
In contrast, the proposed method can restore HS images preserving both details and spectral information without artifacts.

\begin{figure}[t]
\begin{center}
 \begin{minipage}[t]{0.24\hsize}
 \includegraphics[bb = 0 0 560 420, width=1.0\hsize]{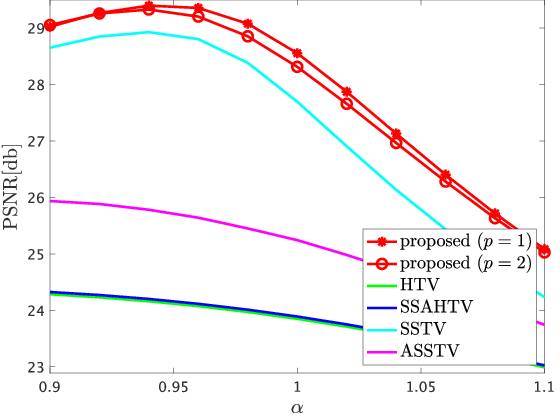}
 \end{minipage}
 \begin{minipage}[t]{0.24\hsize}
 \includegraphics[bb = 0 0 560 420, width=1.0\hsize]{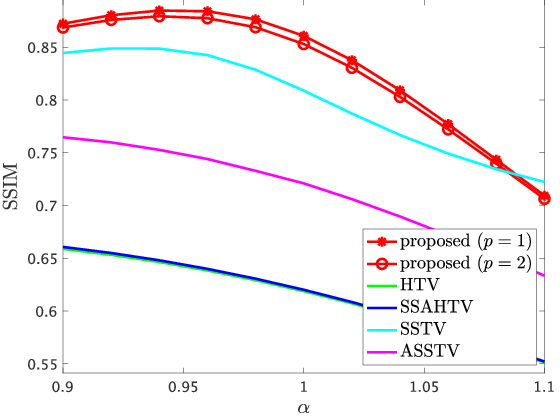}
 \end{minipage}
 \begin{minipage}[t]{0.24\hsize}
 \includegraphics[bb = 0 0 560 420, width=1.0\hsize]{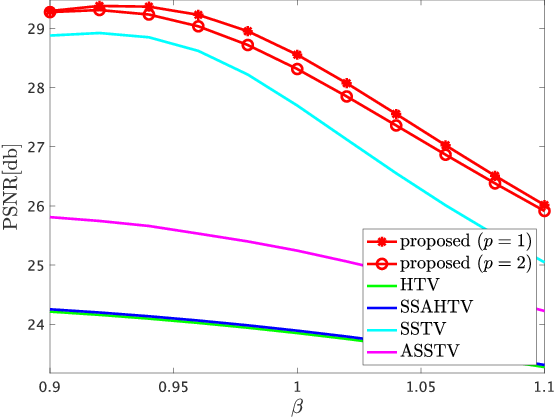}
 \end{minipage}
 \begin{minipage}[t]{0.24\hsize}
 \includegraphics[bb = 0 0 560 420, width=1.0\hsize]{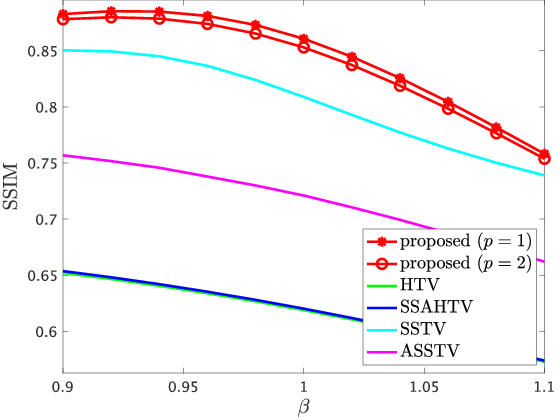}
  \end{minipage}
  
  \begin{minipage}[t]{0.24\hsize}
 \includegraphics[bb = 0 0 560 420, width=1.0\hsize]{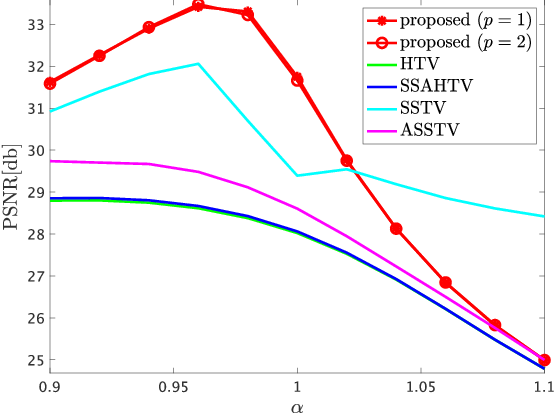}
 \end{minipage}
 \begin{minipage}[t]{0.24\hsize}
 \includegraphics[bb = 0 0 560 420, width=1.0\hsize]{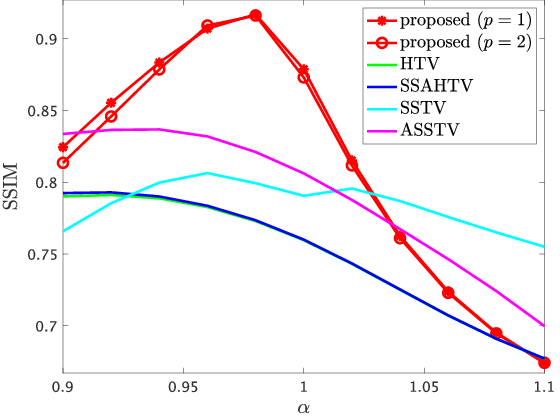}
 \end{minipage}
 \begin{minipage}[t]{0.24\hsize}
 \includegraphics[bb = 0 0 560 420, width=1.0\hsize]{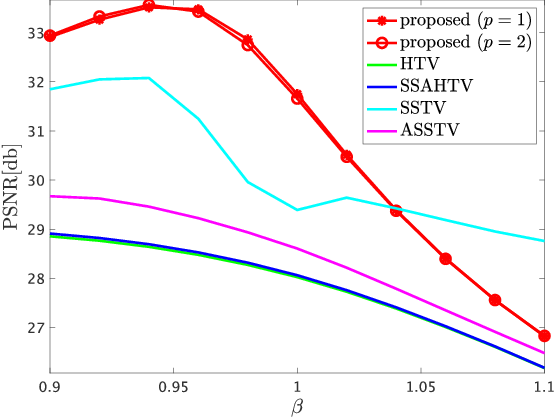}
 \end{minipage}
 \begin{minipage}[t]{0.24\hsize}
 \includegraphics[bb = 0 0 560 420, width=1.0\hsize]{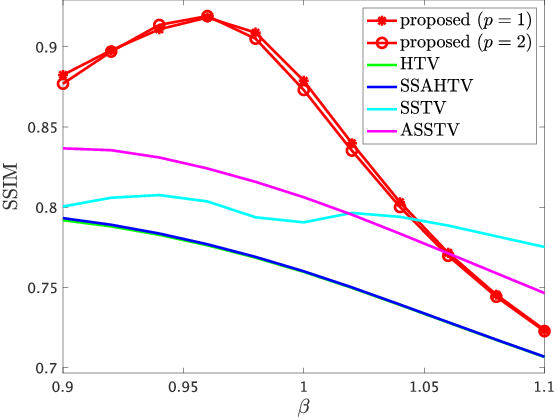}
  \end{minipage}
 
 \begin{minipage}[t]{0.24\hsize}
 \centerline{\footnotesize{PSNR vs $\alpha$}}
 \end{minipage}
 \begin{minipage}[t]{0.24\hsize}
 \centerline{\footnotesize{SSIM vs $\alpha$}}
 \end{minipage}
 \begin{minipage}[t]{0.24\hsize}
 \centerline{\footnotesize{PSNR vs $\beta$}}
 \end{minipage}
 \begin{minipage}[t]{0.24\hsize}
 \centerline{\footnotesize{SSIM vs $\beta$}}
 \end{minipage}
 \caption{ PSNR or SSIM versus $\alpha$ or $\beta$ on the mixed noise removal experiment (the noise level (ii), top: DC, bottom: KSC).}
 \label{parameter_sensitivity}
\end{center}
\end{figure}

Fig.~\ref{omega_graph_mixed} plots PSNR or SSIM of the results by the proposed method versus various $\omega$ in \eqref{HSSTV} changed from $0.01$ to $0.2$, where the values of PSNR and SSIM are averaged over the 13 HS images.
One can see that $\omega \in [0.03,~0.07]$ is a good choice. 
ASSTV and LRTV require to adjust the weight $\tau_b$ and the hyperparameter $\tau$ newly for difference noise intensity, respectively, but the suitable parameter $\omega$ in HSSTV is noise-robust.

Fig.~\ref{other_graph_Gaussian} plots bandwise PSNR and SSIM (left) and spatial and spectral responses (right) of the denoised \textit{Suwannee} HS image in the case of the noise level (ii).
The graphs regarding bandwise PSNR and SSIM show that the proposed method achieves higher-quality restoration than HTV, SSAHTV, and SSTV for all bands and ASSTV and LRMR for most bands.
Besides, even though the proposed method only utilizes HSSTV, the results by the proposed method outperform those by LRTV for some bands in the SSIM cases.
The graph (c) plots the spatial response of the 243rd row of the 30th band.
In the same way, the graph (d) plots the spectral response of the 243rd row and 107th col. 
We can see that the spatial response of the results by HTV and SSAHTV is too smooth compared with the true one.
On the other hand, there exist undesirable variations in the spatial response of the result by SSTV and LRMR.
In contrast, ASSTV, LRTV, and the proposed method restore similar responses to the true one.
In the graph (d), one can see that (i) HTV, SSAHTV, and LRMR produce spectral artifacts, (ii) the shape of the spectral responses of the results by SSTV is similar to the that of the true one, but the mean value is larger than the true one, 
(iii) the spectral response of the results by ASSTV is too smooth and different from the true one,
and (iv) LRTV and the proposed method can restore a spectral response very similar to the true one.

To verify the sensitivity of the parameter $\varepsilon$ and $\eta$, we conducted additional experiments, where we examined various values of $\varepsilon$ and $\eta$. 
Specifically, we set $\varepsilon = 0.83 \alpha \sqrt{NB(1-(s_p(1-l_v-l_h) + l_v + l_h - l_v l_h))\sigma^2}$ and $\eta = \beta NB(0.45 s_p + (l_v + l_h)v_{ave} - l_v l_h v_{ave})$, which are hand-optimized values of the parameters, and changed $\alpha$ and $\beta$ from $0.9$ to $1.1$ at $0.02$ interval (the \textit{DC} and the \textit{KSC} images and the noise level (ii)). 
 Fig.~\ref{parameter_sensitivity} plots PSNR or SSIM of the results by HTV, SSAHTV, SSTV, ASSTV, and the proposed method versus $\alpha$ or $\beta$.
For HTV, SSAHTV, ASSTV, and the proposed method, the graphs show that the suitable values of $\alpha$ and $\beta$ do not vary significantly for both image, and so the parameters $\varepsilon$ and $\eta$ are independent of both a regularization technique and an observed image.
In the SSTV cases, the shapes of the plots are different between KSC and DC.
This is because in the DC case, SSTV converges for all parameter settings, while in the case of KSC, it does not converge when $\alpha, \beta > 1$.

\begin{figure}[t]
\begin{center}
 \begin{minipage}[t]{0.105\hsize}
\includegraphics[bb = 0 0 256 256, width=\hsize]{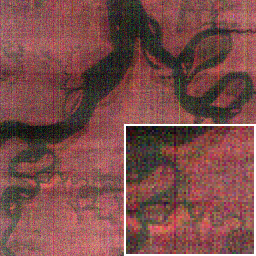}
\end{minipage}
 \begin{minipage}[t]{0.105\hsize}
\includegraphics[bb = 0 0 256 256, width=\hsize]{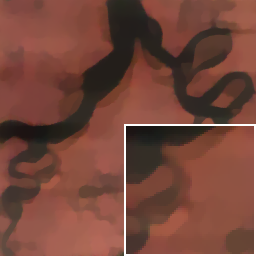}
\end{minipage}
\begin{minipage}[t]{0.105\hsize}
\includegraphics[bb = 0 0 256 256, width=\hsize]{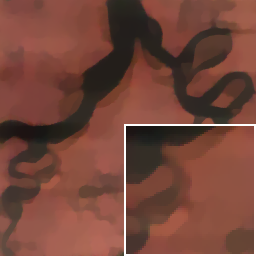}
\end{minipage}
 \begin{minipage}[t]{0.105\hsize}
\includegraphics[bb = 0 0 256 256, width=\hsize]{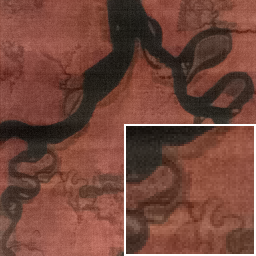}
\end{minipage}
 \begin{minipage}[t]{0.105\hsize}
\includegraphics[bb = 0 0 256 256, width=\hsize]{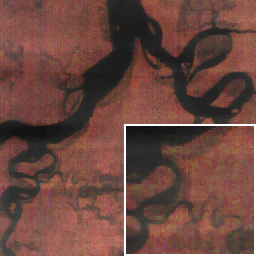}
\end{minipage}
\begin{minipage}[t]{0.105\hsize}
\includegraphics[bb = 0 0 256 256, width=\hsize]{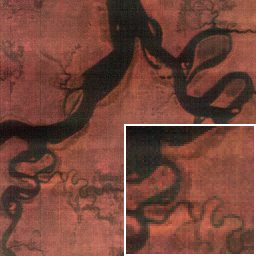}
\end{minipage}
\begin{minipage}[t]{0.105\hsize}
\includegraphics[bb = 0 0 256 256, width=\hsize]{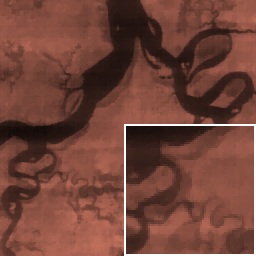}
\end{minipage}
 \begin{minipage}[t]{0.105\hsize}
\includegraphics[bb = 0 0 256 256, width=\hsize]{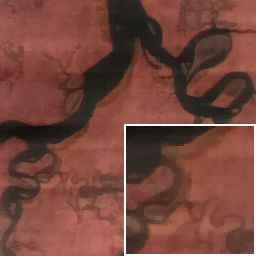}
\end{minipage}
 \begin{minipage}[t]{0.105\hsize}
\includegraphics[bb = 0 0 256 256, width=\hsize]{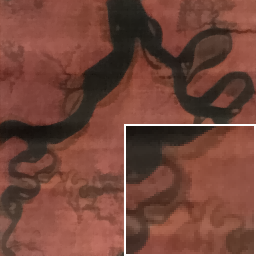}
\end{minipage}

 \begin{minipage}[t]{0.105\hsize}
\centerline{\footnotesize{ observation }}
\end{minipage}
 \begin{minipage}[t]{0.105\hsize}
\centerline{\footnotesize{ HTV }}
\end{minipage}
\begin{minipage}[t]{0.105\hsize}
\centerline{\footnotesize{ SSAHTV }}
\end{minipage}
 \begin{minipage}[t]{0.105\hsize}
\centerline{\footnotesize{ SSTV }}
\end{minipage}
 \begin{minipage}[t]{0.105\hsize}
\centerline{\footnotesize{ ASSTV }}
\end{minipage}
\begin{minipage}[t]{0.105\hsize}
\centerline{\footnotesize{ LRMR }}
\end{minipage}
\begin{minipage}[t]{0.105\hsize}
\centerline{\footnotesize{ LRTV }}
\end{minipage}
 \begin{minipage}[t]{0.105\hsize}
\centerline{\footnotesize{ {\bf{proposed}} }}
\centerline{\footnotesize{ {\bf{($p = 1$)}} }}
\end{minipage}
 \begin{minipage}[t]{0.105\hsize}
\centerline{\footnotesize{ {\bf{proposed}} }}
\centerline{\footnotesize{ {\bf{($p = 2$)}} }}
\end{minipage}

 \caption{The denoising results on the real noise removal experiments.}
 \label{fig:img_real}
\end{center}
\end{figure}

\subsection{Real Noise Removal}
We also examined HTV, SSAHTV, SSTV, ASSTV, LRMR, LRTV, and the proposed method on an HS image with real noise. 
We selected noisy 16 bands from \textit{Suwannee} and used it as a real observed HS image $\v$. 
To maximize the performance of each method, we searched for suitable values of $\sigma$, $s_p$, $l_v$, and $l_h$, and we set the parameters as with Sec. V. A.
Specifically, we set the parameter $\varepsilon = 3.1893$ and $\eta = 31893$ for all TVs.
Besides, 
the parameters $\tau_v$, $\tau_h$, and $\tau_b$ in ASSTV are set as $1$, $1$, $3$, respectively, $r$ and $k$ in LRMR are set as $3$ and $0.1204$, respectively, and $r$ and $tau$ in LRTV are set as $2$ and $0.008$.

Fig~\ref{fig:img_real} shows the results, where the HS images are depicted as RGB images (R = 2nd, G = 6th and B = 13rd bands). 
The results by HTV and SSAHTV have spatial oversmoothing, and SSTV, ASSTV, and LRMR produce spatial artifacts.
Besides, one can see that the results by LRMR and LRTV have spectral artifacts.
On the other hand, the proposed method can restore a detail-preserved HS image without artifacts.

\subsection{Compressed Sensing Reconstruction}\label{subsec:ex_CS}

\begin{table*}[t]
\begin{center}
 \caption{PSNR (left) and SSIM (right) in the CS reconstruction experiment.}
 \label{CS_data}
 \scalebox{0.73}{
 \begin{tabular}{|c|c||c|c|c|c|m{13mm}|m{13mm}||c|c|c|c|m{13mm}|m{13mm}|} \hline
 & & \multicolumn{6}{c||}{PSNR} & \multicolumn{6}{c|}{SSIM} \\ \cline{3-14}
  & $m$ & HTV & SSAHTV & SSTV & ASSTV & {\bf{proposed}} {\bf{($p = 1$)}} & {\bf{proposed}} {\bf{($p = 2$)}} & HTV & SSAHTV & SSTV & ASSTV & {\bf{proposed}} {\bf{($p = 1$)}} & {\bf{proposed}} {\bf{($p = 2$)}} \\ \cline{1-14}
 & 0.4 & 27.46 & 27.49 & 27.53 & 26.51 & \multicolumn{1}{c|}{{\bf{31.15}}} & \multicolumn{1}{c||}{30.71} & 0.6829 & 0.6940 & 0.6013 & 0.6836 & \multicolumn{1}{c|}{{\bf{0.8105}}} & \multicolumn{1}{c|}{0.7948} \\ 
 Beltsville & 0.2 & 26.23 & 26.25 & 24.34 & 24.12 & \multicolumn{1}{c|}{{\bf{29.63}}} & \multicolumn{1}{c||}{29.18} & 0.6363 & 0.6493 & 0.4348 & 0.6108 & \multicolumn{1}{c|}{{\bf{0.7604}}} & \multicolumn{1}{c|}{0.7427} \\ \cline{1-14}
 & 0.4 & 27.97 & 28.02 & 28.49 & 27.68 & \multicolumn{1}{c|}{32.98} & \multicolumn{1}{c||}{{\bf{33.04}}} & 0.7332 & 0.7497 & 0.7377 & 0.7367 & \multicolumn{1}{c|}{0.8902} & \multicolumn{1}{c|}{{\bf{0.8909}}} \\ 
 Suwannee & 0.2 & 26.47 & 26.50 & 25.69 & 25.39 & \multicolumn{1}{c|}{31.37} & \multicolumn{1}{c||}{{\bf{31.44}}} & 0.6810 & 0.7007 & 0.5739 & 0.6633 & \multicolumn{1}{c|}{0.8531} & \multicolumn{1}{c|}{{\bf{0.8534}}} \\ \cline{1-14}
 & 0.4 & 24.69 & 24.73 & 27.33 & 24.71 & \multicolumn{1}{c|}{{\bf{29.70}}} & \multicolumn{1}{c||}{29.29} & 0.6096 & 0.6242 & 0.7522 & 0.6245 & \multicolumn{1}{c|}{{\bf{0.8577}}} & \multicolumn{1}{c|}{0.8460} \\ 
 DC & 0.2 & 23.31 & 23.33 & 24.16 & 22.69 & \multicolumn{1}{c|}{{\bf{27.98}}} & \multicolumn{1}{c||}{27.59} & 0.5215 & 0.5384 & 0.6120 & 0.5037 & \multicolumn{1}{c|}{{\bf{0.7970}}} & \multicolumn{1}{c|}{0.7846} \\ \cline{1-14}
 & 0.4 & 29.94 & 29.96 & 28.21 & 28.59 & \multicolumn{1}{c|}{{\bf{34.36}}} & \multicolumn{1}{c||}{34.34} & 0.7665 & 0.7804 & 0.6826 & 0.7652 & \multicolumn{1}{c|}{0.8882} & \multicolumn{1}{c|}{{\bf{0.8895}}} \\ 
 Cuprite & 0.2 & 28.77 & 28.77 & 25.79 & 26.38 & \multicolumn{1}{c|}{{\bf{32.97}}} & \multicolumn{1}{c||}{32.95} & 0.7368 & 0.7525 & 0.5057 & 0.7207 & \multicolumn{1}{c|}{0.8568} & \multicolumn{1}{c|}{{\bf{0.8578}}} \\ \cline{1-14}
 & 0.4 & 26.99 & 27.05 & 27.82 & 26.49 & \multicolumn{1}{c|}{{\bf{31.80}}} & \multicolumn{1}{c||}{31.61} & 0.6769 & 0.6868 & 0.7414 & 0.6730 & \multicolumn{1}{c|}{{\bf{0.8733}}} & \multicolumn{1}{c|}{0.8705} \\ 
 Reno & 0.2 & 25.57 & 25.61 & 25.57 & 24.52 & \multicolumn{1}{c|}{{\bf{30.22}}} & \multicolumn{1}{c||}{30.04} & 0.6202 & 0.6326 & 0.6276 & 0.5940 & \multicolumn{1}{c|}{{\bf{0.8263}}} & \multicolumn{1}{c|}{0.8228} \\ \cline{1-14}
 & 0.4 & 26.10 & 26.15 & 27.81 & 25.13 & \multicolumn{1}{c|}{{\bf{30.32}}} & \multicolumn{1}{c||}{30.15} & 0.6683 & 0.6803 & 0.7551 & 0.6460 & \multicolumn{1}{c|}{0.8563} & \multicolumn{1}{c|}{{\bf{0.8598}}} \\ 
 Botswana & 0.2 & 24.66 & 24.69 & 24.79 & 22.86 & \multicolumn{1}{c|}{{\bf{28.79}}} & \multicolumn{1}{c||}{28.63} & 0.6014 & 0.6162 & 0.6225 & 0.5519 & \multicolumn{1}{c|}{0.8119} & \multicolumn{1}{c|}{{\bf{0.8163}}} \\ \cline{1-14}
 & 0.4 & 30.54 & 30.55 & 27.65 & 29.55 & \multicolumn{1}{c|}{{\bf{31.36}}} & \multicolumn{1}{c||}{31.04} & 0.7497 & 0.7777 & 0.5066 & 0.7617 & \multicolumn{1}{c|}{{\bf{0.7806}}} & \multicolumn{1}{c|}{0.7655} \\ 
 IndianPines & 0.2 & 29.99 & 29.99 & 25.11 & 28.19 & \multicolumn{1}{c|}{{\bf{30.71}}} & \multicolumn{1}{c||}{30.46} & 0.7366 & {\bf{0.7658}} & 0.3488 & 0.7465 & \multicolumn{1}{c|}{0.7589} & \multicolumn{1}{c|}{0.7491} \\ \cline{1-14}
 & 0.4 & 29.30 & 29.33 & 28.34 & 28.63 & \multicolumn{1}{c|}{{\bf{34.10}}} & \multicolumn{1}{c||}{34.03} & 0.7660 & 0.7742 & 0.6814 & 0.7544 & \multicolumn{1}{c|}{{\bf{0.9019}}} & \multicolumn{1}{c|}{0.9002} \\ 
 KSC & 0.2 & 28.11 & 28.12 & 27.00 & 26.59 & \multicolumn{1}{c|}{{\bf{32.67}}} & \multicolumn{1}{c||}{32.60} & 0.7318 & 0.7410 & 0.6008 & 0.7032 & \multicolumn{1}{c|}{{\bf{0.8698}}} & \multicolumn{1}{c|}{0.8679} \\ \cline{1-14}
& 0.4 & 25.66 & 25.69 & 29.66 & 25.41 & \multicolumn{1}{c|}{{\bf{31.96}}} & \multicolumn{1}{c||}{31.83} & 0.6082 & 0.6205 & 0.8386 & 0.5932 & \multicolumn{1}{c|}{{\bf{0.8900}}} & \multicolumn{1}{c|}{0.8857} \\ 
 PaviaLeft & 0.2 & 24.26 & 24.27 & 27.17 & 23.24 & \multicolumn{1}{c|}{{\bf{30.20}}} & \multicolumn{1}{c||}{30.08} & 0.5103 & 0.5251 & 0.7319 & 0.4434 & \multicolumn{1}{c|}{{\bf{0.8418}}} & \multicolumn{1}{c|}{0.8364} \\ \cline{1-14}
 & 0.4 & 25.83 & 25.85 & 29.86 & 25.61 & \multicolumn{1}{c|}{{\bf{32.45}}} & \multicolumn{1}{c||}{32.21} & 0.6357 & 0.6423 & 0.7962 & 0.6275 & \multicolumn{1}{c|}{{\bf{0.8937}}} & \multicolumn{1}{c|}{0.8877} \\ 
 PaviaRight & 0.2 & 24.30 & 24.30 & 27.54 & 23.61 & \multicolumn{1}{c|}{{\bf{30.56}}} & \multicolumn{1}{c||}{30.38} & 0.5502 & 0.5584 & 0.6917 & 0.5069 & \multicolumn{1}{c|}{{\bf{0.8475}}} & \multicolumn{1}{c|}{0.8392} \\ \cline{1-14}
& 0.4 & 26.49 & 26.53 & 30.02 & 26.38 & \multicolumn{1}{c|}{{\bf{32.88}}} & \multicolumn{1}{c||}{32.70} & 0.6867 & 0.6956 & 0.7830 & 0.6818 & \multicolumn{1}{c|}{{\bf{0.8950}}} & \multicolumn{1}{c|}{0.8901} \\ 
 PaviaU & 0.2 & 24.95 & 24.97 & 27.35 & 24.08 & \multicolumn{1}{c|}{{\bf{31.13}}} & \multicolumn{1}{c||}{30.96} & 0.6138 & 0.6242 & 0.6623 & 0.5680 & \multicolumn{1}{c|}{{\bf{0.8557}}} & \multicolumn{1}{c|}{0.8508} \\ \cline{1-14}
 & 0.4 & 31.19 & 31.24 & 27.69 & 30.18 & \multicolumn{1}{c|}{35.43} & \multicolumn{1}{c||}{{\bf{35.51}}} & 0.8577 & 0.8672 & 0.6153 & 0.8566 & \multicolumn{1}{c|}{0.9222} & \multicolumn{1}{c|}{{\bf{0.9245}}} \\ 
 Salinas & 0.2 & 29.94 & 29.98 & 25.28 & 28.09 & \multicolumn{1}{c|}{34.05} & \multicolumn{1}{c||}{{\bf{34.10}}} & 0.8404 & 0.8516 & 0.4620 & 0.8302 & \multicolumn{1}{c|}{0.9052} & \multicolumn{1}{c|}{{\bf{0.9080}}} \\ \cline{1-14}
 & 0.4 & 30.67 & 30.82 & 27.93 & 28.19 & \multicolumn{1}{c|}{{\bf{34.45}}} & \multicolumn{1}{c||}{34.14} & 0.8647 & 0.8871 & 0.6595 & 0.8489 & \multicolumn{1}{c|}{0.9178} & \multicolumn{1}{c|}{{\bf{0.9208}}} \\ 
 SalinasA & 0.2 & 28.68 & 28.75 & 24.15 & 24.94 & \multicolumn{1}{c|}{{\bf{32.71}}} & \multicolumn{1}{c||}{32.36} & 0.8387 & 0.8655 & 0.4810 & 0.8005 & \multicolumn{1}{c|}{0.8966} & \multicolumn{1}{c|}{{\bf{0.9002}}} \\ \cline{1-14}
 \end{tabular}}
\end{center}
\end{table*}

\begin{figure}[t]
\begin{center}
 \begin{minipage}[t]{0.119\hsize}
\includegraphics[bb = 0 0 256 256, width=\hsize]{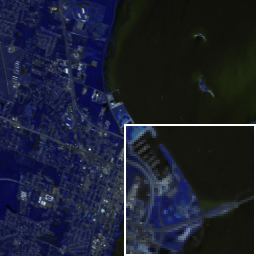}
\end{minipage}
 \begin{minipage}[t]{0.119\hsize}
\includegraphics[bb = 0 0 256 256, width=\hsize]{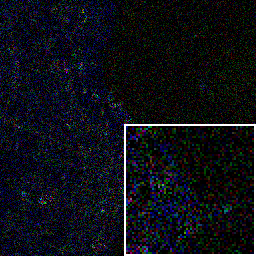}
\end{minipage}
 \begin{minipage}[t]{0.119\hsize}
\includegraphics[bb = 0 0 256 256, width=\hsize]{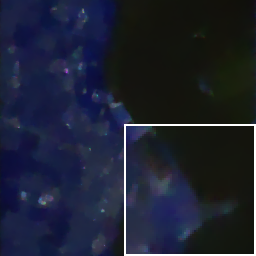}
\end{minipage}
\begin{minipage}[t]{0.119\hsize}
\includegraphics[bb = 0 0 256 256, width=\hsize]{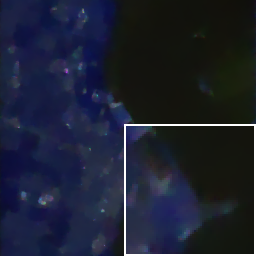}
\end{minipage}
 \begin{minipage}[t]{0.119\hsize}
\includegraphics[bb = 0 0 256 256, width=\hsize]{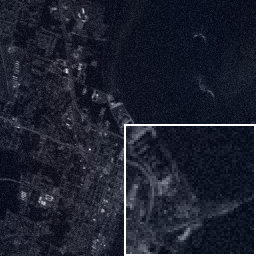}
\end{minipage}
 \begin{minipage}[t]{0.119\hsize}
\includegraphics[bb = 0 0 256 256, width=\hsize]{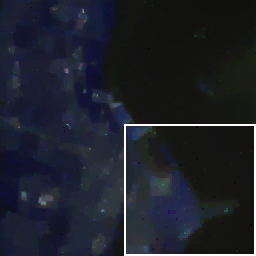}
\end{minipage}
 \begin{minipage}[t]{0.119\hsize}
\includegraphics[bb = 0 0 256 256, width=\hsize]{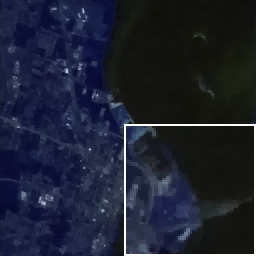}
\end{minipage}
 \begin{minipage}[t]{0.119\hsize}
\includegraphics[bb = 0 0 256 256, width=\hsize]{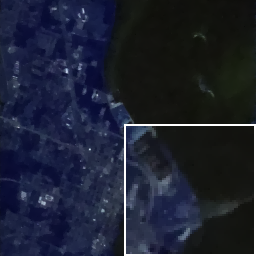}
\end{minipage}

\vspace{-3pt}
 \begin{minipage}[t]{0.119\hsize}
\centerline{\footnotesize{ KSC }}
\end{minipage}
 \begin{minipage}[t]{0.119\hsize}
\centerline{\footnotesize{ 18.78, 0.1267 }}
\end{minipage}
 \begin{minipage}[t]{0.119\hsize}
\centerline{\footnotesize{ 29.30, 0.7660 }}
\end{minipage}
\begin{minipage}[t]{0.119\hsize}
\centerline{\footnotesize{ 29.33, 0.7742 }}
\end{minipage}
 \begin{minipage}[t]{0.119\hsize}
\centerline{\footnotesize{ 28.34, 0.6814 }}
\end{minipage}
 \begin{minipage}[t]{0.119\hsize}
\centerline{\footnotesize{ 28.63, 0.7544 }}
\end{minipage}
 \begin{minipage}[t]{0.119\hsize}
\centerline{\footnotesize{ {\bf{34.10}}, {\bf{0.9019}} }}
\end{minipage}
 \begin{minipage}[t]{0.119\hsize}
\centerline{\footnotesize{ 34.03, 0.9002 }}
\end{minipage}

\vspace{4pt}
 \begin{minipage}[t]{0.119\hsize}
\includegraphics[bb = 0 0 256 256, width=\hsize]{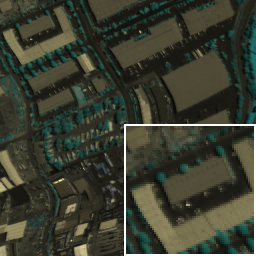}
\end{minipage}
 \begin{minipage}[t]{0.119\hsize}
\includegraphics[bb = 0 0 256 256, width=\hsize]{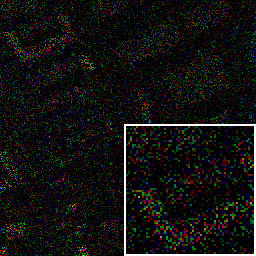}
\end{minipage}
 \begin{minipage}[t]{0.119\hsize}
\includegraphics[bb = 0 0 256 256, width=\hsize]{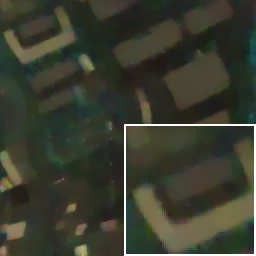}
\end{minipage}
\begin{minipage}[t]{0.119\hsize}
\includegraphics[bb = 0 0 256 256, width=\hsize]{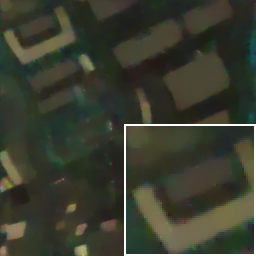}
\end{minipage}
 \begin{minipage}[t]{0.119\hsize}
\includegraphics[bb = 0 0 256 256, width=\hsize]{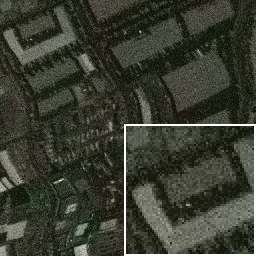}
\end{minipage}
 \begin{minipage}[t]{0.119\hsize}
\includegraphics[bb = 0 0 256 256, width=\hsize]{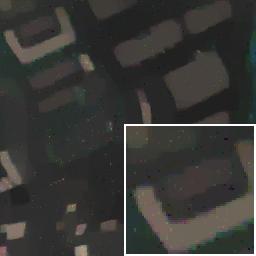}
\end{minipage}
 \begin{minipage}[t]{0.119\hsize}
\includegraphics[bb = 0 0 256 256, width=\hsize]{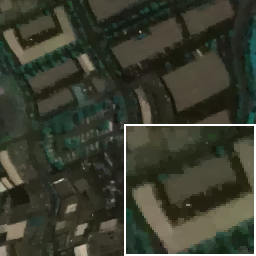}
\end{minipage}
 \begin{minipage}[t]{0.119\hsize}
\includegraphics[bb = 0 0 256 256, width=\hsize]{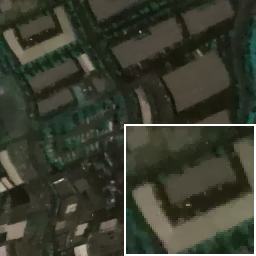}
\end{minipage}

\vspace{-3pt}
 \begin{minipage}[t]{0.119\hsize}
\centerline{\footnotesize{ Reno }}
\end{minipage}
 \begin{minipage}[t]{0.119\hsize}
\centerline{\footnotesize{ 14.11, 0.07624}}
\end{minipage}
 \begin{minipage}[t]{0.119\hsize}
\centerline{\footnotesize{ 25.57, 0.6202 }}
\end{minipage}
 \begin{minipage}[t]{0.119\hsize}
\centerline{\footnotesize{ 25.61, 0.6326 }}
\end{minipage}
 \begin{minipage}[t]{0.119\hsize}
\centerline{\footnotesize{ 25.57, 0.6276 }}
\end{minipage}
 \begin{minipage}[t]{0.119\hsize}
\centerline{\footnotesize{ 24.52, 0.5940 }}
\end{minipage}
 \begin{minipage}[t]{0.119\hsize}
\centerline{\footnotesize{ {\bf{30.22}}, {\bf{0.8263}} }}
\end{minipage}
 \begin{minipage}[t]{0.119\hsize}
\centerline{\footnotesize{ 30.04, 0.8228 }}
\end{minipage}

 \begin{minipage}[t]{0.119\hsize}
\centerline{\footnotesize{ groundtruth }}
\end{minipage}
 \begin{minipage}[t]{0.119\hsize}
\centerline{\footnotesize{ observation }}
\end{minipage}
 \begin{minipage}[t]{0.119\hsize}
\centerline{\footnotesize{ HTV }}
\end{minipage}
 \begin{minipage}[t]{0.119\hsize}
\centerline{\footnotesize{ SSAHTV }}
\end{minipage}
 \begin{minipage}[t]{0.119\hsize}
\centerline{\footnotesize{ SSTV }}
\end{minipage}
 \begin{minipage}[t]{0.119\hsize}
\centerline{\footnotesize{ ASSTV }}
\end{minipage}
 \begin{minipage}[t]{0.119\hsize}
\centerline{\footnotesize{ \bf{proposed}}}
\centerline{\footnotesize{\bf{($p = 1$)}}}
\end{minipage}
 \begin{minipage}[t]{0.119\hsize}
\centerline{\footnotesize{ \bf{proposed}}}
\centerline{\footnotesize{\bf{($p = 2$)}}}
\end{minipage}

 \caption{Resulting HS images with their PSNR (left) and SSIM (right) on the CS reconstruction experiment (top: KSC, $m = 0.4$, bottom: Reno, $m = 0.2$).}
 \label{fig:img_CS}
\end{center}
\end{figure}

\begin{figure}[t]
\begin{center}
 \begin{minipage}[t]{0.24\hsize}
 \includegraphics[bb = 0 0 560 420, width=1.0\hsize]{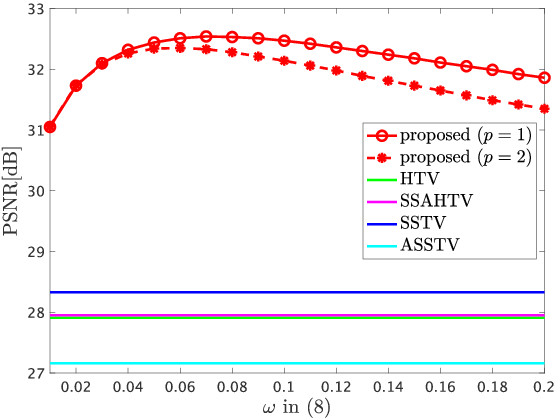}
 \centerline{\footnotesize{PSNR, $m = 0.4$}}
 \end{minipage}
 \begin{minipage}[t]{0.24\hsize}
 \includegraphics[bb = 0 0 560 420, width=1.0\hsize]{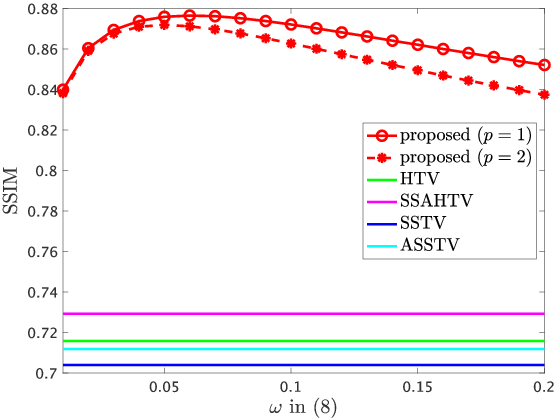}
 \centerline{\footnotesize{SSIM, $m = 0.4$}}
 \end{minipage}
 \begin{minipage}[t]{0.24\hsize}
 \includegraphics[bb = 0 0 560 420, width=1.0\hsize]{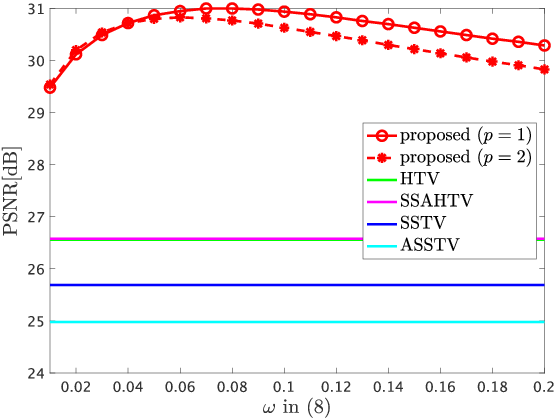}
 \centerline{\footnotesize{PSNR, $m = 0.2$}}
 \end{minipage}
 \begin{minipage}[t]{0.24\hsize}
 \includegraphics[bb = 0 0 560 420, width=1.0\hsize]{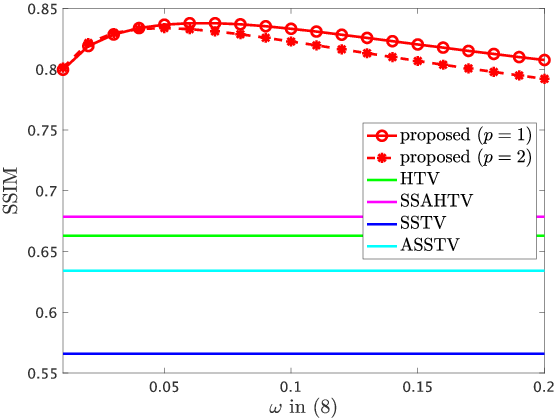}
 \centerline{\footnotesize{SSIM, $m = 0.2$}}
 \end{minipage}
 \caption{PSNR or SSIM versus $\omega$ in \eqref{HSSTV} on the CS reconstruction experiment.} 
 \label{omega_graph_CS}
\end{center}
\end{figure}

\begin{figure}[t]
\begin{center}
 \begin{minipage}[t]{0.24\hsize}
 \includegraphics[bb = 0 0 560 420, width=1.0\hsize]{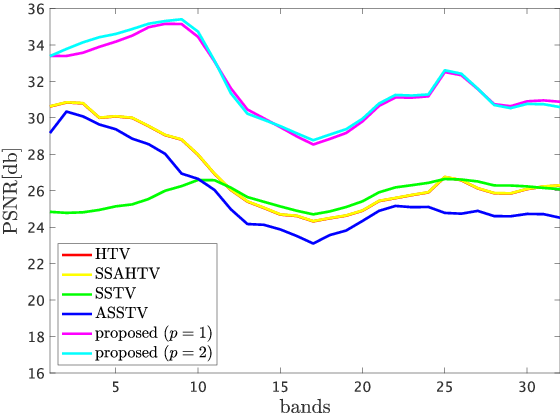}
 \centerline{\footnotesize{(a) Bandwise PSNR}}
 \end{minipage}
 \begin{minipage}[t]{0.24\hsize}
 \includegraphics[bb = 0 0 560 420, width=1.0\hsize]{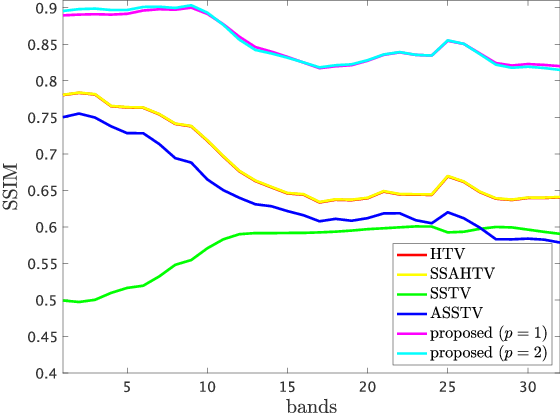}
 \centerline{\footnotesize{(b) Bandwise SSIM}}
 \end{minipage}
 \begin{minipage}[t]{0.24\hsize}
 \includegraphics[bb = 0 0 560 420, width=1.0\hsize]{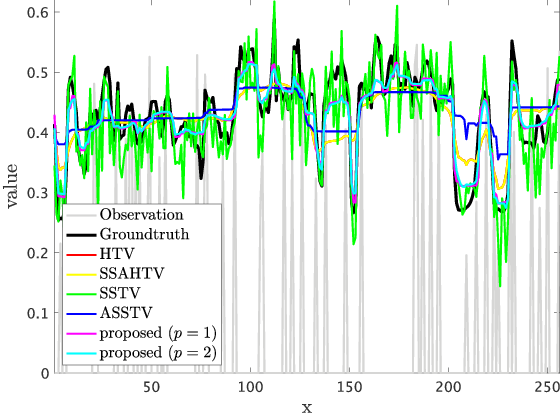}
 \centerline{\footnotesize{(c) Spatial response}}
 \end{minipage}
 \begin{minipage}[t]{0.24\hsize}
 \includegraphics[bb = 0 0 560 420, width=1.0\hsize]{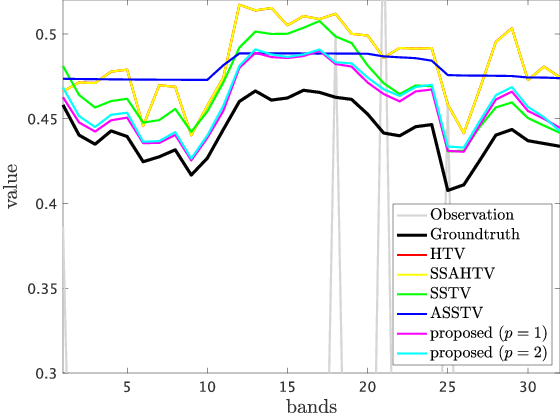}
 \centerline{\footnotesize{(d) Spectral response}}
 \end{minipage}
 \caption{ Bandwise PSNR and SSIM and spatial and spectral responses on the CS reconstruction experiment (Suwannee).}
 \label{other_graph_CS}
\end{center}
\end{figure}

We conducted on an experiment on compressed sensing (CS) reconstruction \cite{CSBaraniuk,CandesCSSPM}.
The CS theory says that high-dimensional signal information can be reconstructed from incomplete random measurements by exploiting sparsity in some domains, e.g., the gradient domain (TV). 
In general, HS imaging captures an HS image by scanning 1D spatial and spectral information, because it senses spectral information by dispersing the incident light. 
Therefore, 
capturing moving objects is very difficult in HS imaging. 
To overcome the drawback, one-shot HS imaging based on CS has been actively studied \cite{one-shotHSI1, one-shotHSI2}.

In this experiment, we assume that 
$\PHI\in\R^{M \times NB}$ in \eqref{model} is a random sampling matrix with the sampling rate $m$ ($0 < m < 1$ and $M = mNB$).
Here, since $\PHI$ is a semi-orthogonal matrix, we can efficiently solve the problem as explained in Sec.~\ref{subsec:restoration}.
Moreover, since the main objective in the experiments is to verify CS reconstruction performance by HSSTV, we assume that the observations are contaminated by only an additive white Gaussian noise $\n$ with noise intensity $\sigma = 0.1$. 

We set the CS reconstruction problem as follows:
\begin{equation*}
\min_{\u} \HSSTV(\u) \mbox{ s.t. } \left[
\begin{array}{l}
\PHI \u \in \Bmath_{2,\varepsilon}^{\v}, \\
\u \in [\mu_{\min}, \mu_{\max}]^{NB}.
\end{array}
\right.
\end{equation*}
The problem is derived by removing the second constraint and $\s$ from Prob.~\eqref{prob:HSSTV_Gaussian}.
Therefore, we can solve the above problem by removing $\s$, $\z_3$, and $\d_3$ in Alg.~\eqref{alg:ADMMHSSTV} and replacing $\z_4$ and $\d_4$ with $\z_3$ and $\d_3$, respectively. 
As in Sec.~\ref{subsec:restoration}, the update of $\u$ is strictly-convex quadratic minimization, and so it comes down to
\begin{align*}
&\u^{(n+1)} = (\A_{\omega}^{\top} \A_{\omega} + \PHI^{\top} \PHI + \I)^{-1} \mbox{RHS}, \\
&\mbox{RHS} = \A_{\omega}(\z_1^{(n)} - \d_1^{(n)}) + \PHI^{\top}(\z_2^{(n)} - \d_2^{(n)}) + (\z_3^{(n)} - \d_3^{(n)}).
\end{align*}
We set $m=0.2$ or $0.4$ and $\varepsilon = \sqrt{mNB\sigma^2}$. 
In the ASSTV case, we set the parameters $(\tau_v, \tau_h, \tau_b) = (1, 1, 0.5)$, which experimentally achieves the best performance.

Tab.~\ref{CS_data} shows PSNR and SSIM of the reconstructed HS images. 
For all $m$ and HS images, both PSNR and SSIM of the results by the proposed method are almost higher than that by HTV, SSAHTV, SSTV, and ASSTV. 

Fig.~\ref{fig:img_CS} is the reconstructed results on \textit{KSC} and \textit{Reno} with the random sampling ratio $m = 0.4$ and $0.2$, respectively.
Here, 
the HS images are depicted as RGB images (R = 8th, G = 16th and B = 32nd bands).
One can see that (i) HTV and SSAHTV cause spatial oversmoothing, (ii) SSTV produces artifacts and spectral distortion, where it appears as the difference from the color of the true HS images, and (iii) the results by ASSTV have spatial oversmoothing and spectral distortion.
On the other hand, the proposed method reconstructs meaningful details without both artifacts and spectral distortion.

Fig.~\ref{omega_graph_CS} plots PSNR or SSIM of the results by the proposed method versus $\omega$ averaged over the 13 HS images for each $m$. 
The graphs show that $\omega\in[0.05,0.1]$ is a good choice in most cases.
In comparison with Fig.~\eqref{omega_graph_mixed}, the suitable range of $\omega$ in CS reconstruction is almost the same as that in denoising.

Fig.~\ref{other_graph_CS} plots bandwise PSNR or SSIM (left) and spatial and spectral responses (right) (\textit{Suwannee}, $m = 0.2$).
According to bandwise PSNR and SSIM, one can see that the proposed method achieves higher-quality reconstruction for all bands than HTV, SSAHTV, SSTV, and ASSTV.
The graphs (c) and (d) plot the spatial and spectral responses of the same position in Sec.~\ref{subsec:ex_denoising}. 
The graph (c) shows that (i) the spatial response of the results by HTV, SSAHTV, and ASSTV are oversmoothing, (ii) SSTV produces undesirable variation, and (iii) the spatial response reconstructed by the proposed method is similar to the true one.
In the graph (d), HTV and SSAHTV generate undesirable variation, and ASSTV causes oversmoothing.
Thanks to the evaluation of spatio-spectral piecewise-smoothness, SSTV reconstructs a similar spectral response to the true one, but the mean values are larger than the true one.
The proposed method achieves the most similar reconstruction of spectral response among all the TVs.

\section{Conclusion}\label{sec:C}
We have proposed a new constrained optimization approach to HS image restoration.
Our proposed method is formulated as a convex optimization problem, where we utilize a novel regularization technique named HSSTV and incorporate data-fidelity as hard constraints.
HSSTV evaluates direct spatial piecewise-smoothness and spatio-spectral piecewise-smoothness, and so it has a strong ability of HS restoration.
Thanks to the design of the constraint-type data-fidelity, we can independently set the hyperparameters that balance between regularization and data-fidelity.
To solve the proposed problem, we develop an efficient algorithm based on ADMM.
Experimental results on mixed noise removal, real noise removal, and CS reconstruction demonstrate the advantages of the proposed method over various HS image restoration methods.

\authorcontributions{Conceptualization, S.T., S.O., and I.K.; methodology, S.T. and S.O.; software, S.T.; validation, S.T.; formal analysis, S.T.; investigation, S.T.; writing—original draft, S.T.; writing—review and editing, S.O. and I.K.; supervision, S.O., and I.K.; project administration, S.T., S.O., and I.K.; funding acquisition, S.T., S.O., and I.K.  All authors read and agreed to the published version of the manuscript.}

\funding{This work was supported in part by JST CREST under Grant JPMJCR1662 and JPMJCR1666, and in part by JSPS KAKENHI under Grant 18J20290, 18H05413, and 20H02145.} 


\conflictsofinterest{The authors declare no conflict of interest.}

%

%

\reftitle{References}


\externalbibliography{yes}
\bibliography{TCI_HSSTV.bib}




\end{document}